\documentclass[iop]{emulateapj}

\newcommand\kms{\ifmmode{\rm km\thinspace s^{-1}}\else km\thinspace s$^{-1}$\fi}
\newcommand\nar{New Astronomy Reviews}

\shortauthors{Torres et al.}
\shorttitle{Atlas}

\begin{document}
\submitted{Accepted for publication in The Astrophysical Journal}

\title{Orbital and Physical properties of the Pleiades Binary 27 Tau (Atlas)}

\author{
Guillermo Torres\altaffilmark{1}, 
Andrew Tkachenko\altaffilmark{2}, 
Kre\v{s}imir Pavlovski\altaffilmark{3}, 
Seth Gossage\altaffilmark{4}, 
Gail H.\ Schaefer\altaffilmark{5}, 
Carl Melis\altaffilmark{6}, 
Michael Ireland\altaffilmark{7}, 
John D.\ Monnier\altaffilmark{8}, 
Narsireddy Anugu\altaffilmark{5}, 
Stefan Kraus\altaffilmark{9}, 
Cyprien Lanthermann\altaffilmark{5}, 
Kathryn Gordon\altaffilmark{10}, 
Robert Klement\altaffilmark{11}, 
Simon J.\ Murphy\altaffilmark{12}, and 
Rachael M.\ Roettenbacher\altaffilmark{8} 
}

\altaffiltext{1}{Center for Astrophysics $\vert$ Harvard \& Smithsonian, 60
  Garden St., Cambridge, MA 02138, USA; gtorres@cfa.harvard.edu}

\altaffiltext{2}{Institute of Astronomy, KU Leuven, Celestijnenlaan 200D, 3001 Leuven, Belgium}

\altaffiltext{3}{Department of Physics, Faculty of Science, University of Zagreb, 10000 Zagreb, Croatia}

\altaffiltext{4}{Center for Interdisciplinary Exploration and Research in Astrophysics (CIERA), Evanston, IL 60201, USA}

\altaffiltext{5}{The CHARA Array of Georgia State University, Mount Wilson Observatory, Mount Wilson, CA 91023, USA}

\altaffiltext{6}{Department of Astronomy \& Astrophysics, University of California, La Jolla, CA 92093-0424, USA}

\altaffiltext{7}{Research School of Astronomy and Astrophysics, Australian National University, Canberra ACT 2601, Australia}

\altaffiltext{8}{Astronomy Department, University of Michigan, Ann Arbor, MI 48109, USA}

\altaffiltext{9}{Astrophysics Group, Department of Physics \& Astronomy, University of Exeter, Exeter EX4 4QL, UK}

\altaffiltext{10}{Mitchell Community College, Statesville, NC 28677, USA}

\altaffiltext{11}{European Organisation for Astronomical Research in the Southern Hemisphere (ESO), Casilla 19001, Santiago 19, Chile}

\altaffiltext{12}{Centre for Astrophysics, University of Southern Queensland, Toowoomba, QLD 4350, Australia}

\begin{abstract}
We report new spectroscopic and interferometric observations of the
Pleiades binary star Atlas, which played an important role nearly
three decades ago in settling the debate over the distance to the
cluster from ground-based and space-based determinations. We use the
new measurements, together with other published and archival
astrometric observations, to improve the determination of the 291-day
orbit and the distance to Atlas ($136.2 \pm 1.4$~pc). We also derive the main
properties of the components, including their absolute masses ($5.04
\pm 0.17~M_{\odot}$ and $3.64 \pm 0.12~M_{\odot}$), sizes, effective
temperatures, projected rotational velocities, and chemical
composition.  We find that the more evolved primary star is
rotationally distorted, and are able to estimate its oblateness
and the approximate {orientation} of its spin axis from
the interferometric observations. {The spin
axis may well be aligned with the orbital axis}.
Models of stellar evolution from MESA that
account for rotation provide a good match to all of the primary's
global properties, and point to an initial angular rotation rate on
the zero-age main sequence of about 55\% of the breakup velocity. The
current location of the star in the H-R diagram is near the very end
of the hydrogen-burning main sequence, at an age of about 105~Myr,
according to these models. Our spectroscopic analysis of the more 
slowly-rotating secondary indicates that it is a helium-weak star, with other chemical anomalies.
\end{abstract}

\section{Introduction}
\label{sec:introduction}

After the publication of the Hipparcos catalog
\citep{ESA:1997}, a heated debate ensued in the literature about the
mean distance to the Pleiades cluster. Significant tension was
found between the result from the satellite observations, $\sim$118\,pc
\citep{vanLeeuwen:1999}, and the classical ground-based
determinations, which consistently pointed to larger distances of
$\sim$130\,pc \citep[e.g.,][]{Meynet:1993, vanLeeuwen:2009a}. For reviews on
this subject, the reader may consult, e.g., \cite{Stello:2001},
\cite{Percival:2005}, or \cite{vanLeeuwen:2007, vanLeeuwen:2009b}.

Binary stars in the cluster featured prominently among the approaches to this
problem, as they permit independent estimates of the distance that can
be very precise. One of the objects used for this purpose was the
A-type eclipsing binary HD~23642 \citep[V1229~Tau;][]{Miles:1999,
  Torres:2003}. It was first analyzed by \cite{Munari:2004} to argue
convincingly in favor of the ground-based distance determinations, and
was later reanalyzed by others \citep{Southworth:2005, Groenewegen:2007,
  David:2016, Southworth:2023}. Another Pleiades binary that played a central role is
Atlas (27~Tau, HD~23850, $V = 3.63$, \ion{B8}{5}), the second
brightest object in the cluster, and the subject of this paper.

Atlas is a 291\,day binary whose duplicity was discovered in 1968 by
the lunar occultation technique \citep{Nather:1970,
  Evans:1971}.\footnote{These authors initially expressed
  reservations about their detection. However, in a subsequent paper
  with others \citep{McGraw:1974}, the observation was
  re-reduced and found to be supported by additional occultation
  events, confirming it as a genuine detection. See below for a much
  earlier but spurious claim of duplicity.} It was also
resolved much later by \cite{Pan:2004}, using long-baseline
interferometry. These authors combined their well constrained
astrometric orbit with a mass-luminosity relation, and obtained an
estimate of the distance that was again consistent with the
traditional ground-based values.  \cite{Zwahlen:2004} improved on
this analysis by providing further interferometric observations, but
more importantly, by obtaining the first double-lined spectroscopic
orbit for Atlas. This permitted a completely
model-independent determination of the distance via the orbital
parallax. 
These two binaries, along with other more direct methods of
measuring the distance that included space-based trigonometric
parallax determinations with HST \citep{Soderblom:2005} and VLBI
parallax measurements from radiointerferometry \citep{Melis:2014},
finally settled the distance debate and demonstrated convincingly that
Hipparcos was in error.

Along with the interferometric orbit of Atlas, the \cite{Zwahlen:2004}
study found that both B-type components are rapid rotators, with $v \sin i$
values they estimated to be about 240~\kms\ for the primary and 60~\kms\ for the secondary.
Their masses were inferred to be 4.74 and 3.42~$M_{\sun}$, with
relative uncertainties of 5.3\% and 7.3\%, respectively. 
Additionally, the angular
diameter of the primary has been measured more recently by
\cite{Gordon:2019}, which enables its absolute radius to be inferred,
once a distance is adopted. All of these properties make Atlas potentially
useful for an independent age determination for the Pleiades, as its
location at the very tip of the main sequence of the cluster means it
is among the most sensitive and favorable objects for that purpose.

Because of this, and in view of the historical importance of Atlas in
the context of the Hipparcos distance controversy, we have 
gathered independent spectroscopic observations to
revisit the determination of its properties, and to compare them for
the first time against current models of stellar evolution. Additional
interferometric observations have been obtained as well, strengthening
the constraint on the size of the primary.

The paper is structured as follows. The new spectroscopic observations
are described in Section~\ref{sec:spectroscopy}. We report our
interferometric observations in Section~\ref{sec:astrometry}, where we
also mention other existing astrometric observations for Atlas from
the literature, including lunar occultation measurements, and the
Hipparcos observations. In Section~\ref{sec:disentangling} we explain
our approach to separating the spectra of the individual components
with the technique of spectral disentangling, which we then use in the
following section to determine the spectroscopic properties of both
stars (effective temperature, surface gravity, projected rotational
velocity, metallicity, etc.). The radial velocities {(RVs)} we infer from our
spectra are described and reported in Section~\ref{sec:rvs}. The
details of our orbital analysis combining the velocities and the
astrometry are presented in Section~\ref{sec:orbit}, along with the
results for Atlas. The physical properties of the two components are
then discussed and compared with current models of stellar evolution in
Section~\ref{sec:discussion}, and in Section~\ref{sec:conclusions} we
draw our conclusions.

\section{Spectroscopic Observations}
\label{sec:spectroscopy}

Atlas was observed spectroscopically at the Center for Astrophysics
(CfA) beginning in October of 2017, using the Tillinghast Reflector
\'Echelle Spectrograph \citep[TRES;][]{Furesz:2008, Szentgyorgyi:2007}
on the 1.5\,m Tillinghast reflector at the Fred L.\ Whipple Observatory
on Mount Hopkins (Arizona, USA). This is a bench-mounted, fiber-fed
instrument providing a resolving power of $R \approx 44,\!000$, with a
CCD detector recording 51 \'echelle orders between 3800--9100~\AA. We
collected 48 spectra through March of 2022, with signal-to-noise
ratios ranging from 190 to 860 per resolution element of 6.8~\kms.
Reductions were performed with a dedicated pipeline
\citep[see][]{Buchhave:2012}, and wavelength solutions relied on
exposures of a thorium-argon lamp taken before and after each science
exposure. The determination of {RVs} from this material is
described later in Section~\ref{sec:rvs}.

\section{Astrometric Observations}
\label{sec:astrometry}

\subsection{CHARA Array}
\label{sec:chara}

The Center for High Angular Resolution Astronomy (CHARA) Array is a long-baseline optical interferometer operated by Georgia State University, and located at Mount Wilson Observatory in southern California (USA). The CHARA Array combines the light from up to six 1\,m telescopes (S1, S2, E1, E2, W1, W2) with baselines ranging from 34 to 331\,m \citep{tenBrummelaar:2005, tenBrummelaar:2016}. We used new and archival observations of Atlas recorded with the MIRC-X \citep{Anugu:2020}, MYSTIC \citep{Setterholm:2023}, CLIMB \citep{tenBrummelaar:2013}, and PAVO \citep{Ireland:2008} beam combiners. Table~\ref{tab:charalog} presents a log of CHARA observations that lists the UT date, instrument, bandpass ($\lambda$), telescopes, number of data sets recorded on Atlas ($N$), and calibrator stars. To calibrate the interferometric transfer function, observations of either unresolved or small angular diameter calibrator stars were interspersed before and after the observations of Atlas. Table~\ref{tab:caldiam} lists the uniform disk diameters (UDD) adopted for the calibrators \citep{Bourges:2014}.

MIRC-X combines the light from six telescopes and operates in the near-infrared $H$ band. In 2022, we also recorded data with the six telescope $K$-band combiner MYSTIC that operates simultaneously with MIRC-X. The MIRC-X and MYSTIC data were recorded in the low spectral resolution mode ($R = \lambda/\Delta\lambda$ = 50 prism) and reduced using the standard pipeline for MIRC-X/MYSTIC\footnote{\url{https://zenodo.org/records/12735292}}
\citep[version 1.3.5 and 1.4.0;][]{Anugu:2020}. The pipeline produces squared visibilities and closure phases for each baseline over 8--10 spectral channels. We used an integration time of 150~sec.

CLIMB combines the light from three telescopes and operates in the near-infrared $H$ and $K$ bands. The CLIMB data were originally published by \citet{Gordon:2019}. We re-reduced the CLIMB data using the pipeline developed by J.\ D.\ Monnier, with the general method described by \citet{Monnier:2011} extended to three beams \citep[e.g.,][]{Kluska:2018}, producing broadband squared visibilities for each baseline and closure phases for each closed triangle.

PAVO is a visible light combiner that operates in the $R$ band (0.65--0.79~$\mu$m). The data on Atlas were collected in two-telescope mode, and reduced with the standard PAVO reduction pipeline\footnote{\url{https://www.chara.gsu.edu/tutorials/pavo-data-reduction}}, producing squared visibilities over $\sim$23 spectral channels.

The calibrated data sets from PAVO, CLIMB, MIRC-X, and MYSTIC will be
made available in the OIFITS format
through the Jean-Marie Mariotti Center (JMMC) Optical Interferometry Database\footnote{\url{https://www.jmmc.fr/english/tools/data-bases/oidb/}} (OIDB).

\setlength{\tabcolsep}{4pt}
\begin{deluxetable*}{lccccl}
\tablewidth{0pc}
\tablecaption{Log of Observations at the CHARA Array \label{tab:charalog}}
\tablehead{
\colhead{UT Date} &
\colhead{Instrument} &
\colhead{$\lambda$} &
\colhead{Telescopes} &
\colhead{$N$} &
\colhead{Calibrators} }
\startdata
2012Nov15 & PAVO   & $R$  & S2E2         & 1  & HD 23753                       \\ 
2014Sep22 & CLIMB  & $H$  & S1E1W1       & 1  & HD 23338                       \\ 
2014Sep23 & CLIMB  & $H$  & S1E1W1       & 2  & HD 23338, HD 23753             \\ 
2014Sep24 & CLIMB  & $H$  & S1E1W1       & 2  & HD 23338, HD 23753             \\ 
2015Sep03 & PAVO   & $R$  & E2W2         & 3  & HD 21050, HD 25175, HD 22860   \\ 
2015Sep05 & PAVO   & $R$  & S1W1         & 3  & HD 21050, HD 25175		    \\ 
2015Sep06 & PAVO   & $R$  & E2S1         & 4  & HD 21050, HD 25175, HD 24368   \\
2015Sep08 & PAVO   & $R$  & W2E1         & 2  & HD 21050, HD 25175		    \\ 
2015Sep09 & PAVO   & $R$  & S2W2         & 1  & HD 22860			    \\ 
2015Sep10 & PAVO   & $R$  & S2W2         & 1  & HD 21050, HD 22860		    \\ 
2015Sep14 & CLIMB  & $H$  & S1E1W1       & 3  & HD 23338, HD 23753             \\ 
2015Nov08 & PAVO   & $R$  & W2S2         & 3  & HD 21050, HD 25175		    \\ 
2017Nov19\tablenotemark{a} & MIRC-X & $H$  & E1W2W1S2S1E2 & 2  &                     	    \\ 
2017Nov20\tablenotemark{b} & MIRC-X & $H$  & E1W2W1S2S1E2 & 1  & HD 32630                	    \\  
2018Dec10 & MIRC-X & $H$  & E1W2W1S2S1E2 & 1  & HD 16730, HD 23232, HD 24398, HD 23183 \\ 
2022Nov15\tablenotemark{c} & MIRC-X & $H$  & E1W2W1E2     & 1  & HD 218235, HD 20150, HD 23288, HD 27627, HD 27808, HD 28406, HD 36667 \\ 
2022Nov15\tablenotemark{c} & MYSTIC & $K$  & E1W2W1E2     & 1  & HD 218235, HD 20150, HD 23288, HD 27627, HD 27808, HD 28406, HD 36667 
\enddata
\tablecomments{Column $N$ is the number of data sets recorded on Atlas on each night.}
\tablenotetext{a}{The observing log for UT 2017 Nov 19 indicates that Atlas was observed as a calibrator. Atlas was observed with all 6 telescopes, while the next nearest calibrators were more than 60$^\circ$ away, observed more than 4 hours later, and obtained fringes on only 3 or 4 telescopes. The Atlas data were not calibrated, so only the closure phases were used in the binary fit.}
\tablenotetext{b}{The observing log for UT 2017 Nov 20 indicates that Atlas was observed as a calibrator. The closest 6-telescope calibrator was located 24$^\circ$ away and observed 4 hours later using a different pop configuration. This resulted in poor calibration of the visibilities ($V^2 > 1$ on several baselines). Only the closure phases were used in the binary fit.}
\tablenotetext{c}{On UT 2022 Nov 15, the S1 and S2 telescopes were offline because of a problem with the metrology signal on the S1 delay line cart, and a mechanical problem with the drive bearings on the S2 telescope.}
\end{deluxetable*}

\setlength{\tabcolsep}{3pt}
\begin{deluxetable}{lccc@{\hskip 1.5pc}lccc}
\tablewidth{0pc}
\tablecaption{Adopted Calibrator Uniform Disk Diameters\label{tab:caldiam}}
\tablehead{
\colhead{Calibrator} &
\colhead{$\lambda$} &
\colhead{UDD} &
\colhead{$\sigma_{\rm UDD}$~~~~~~} &
\colhead{Calibrator} &
\colhead{$\lambda$} &
\colhead{UDD} &
\colhead{$\sigma_{\rm UDD}$} \\
\colhead{} &
\colhead{} &
\colhead{(mas)} &
\colhead{(mas)~~~~~~} &
\colhead{} &
\colhead{} &
\colhead{(mas)} &
\colhead{(mas)}
}
\startdata
HD 16730   & $H$  & 0.656  & 0.015 &                  HD 24368   & $R$  & 0.216  & 0.006 \\                   
HD 20150   & $H$  & 0.350  & 0.013 &                  HD 24398\tablenotemark{b}   & $H$  & 0.636  & 0.026 \\  
HD 20150   & $K$  & 0.350  & 0.013 &                  HD 25175   & $R$  & 0.193  & 0.005 \\                   
HD 21050   & $R$  & 0.187  & 0.005 &                  HD 27627   & $H$  & 0.273  & 0.006 \\                   
HD 22860\tablenotemark{a}   & $R$  & 0.141  & 0.008 & HD 27627   & $K$  & 0.273  & 0.006 \\                   
HD 23183   & $H$  & 0.829  & 0.064 &                  HD 27808   & $H$  & 0.275  & 0.007 \\                   
HD 23232   & $H$  & 0.653  & 0.016 &                  HD 27808   & $K$  & 0.275  & 0.007 \\                   
HD 23288   & $H$  & 0.228  & 0.007 &                  HD 28406   & $H$  & 0.277  & 0.007 \\                   
HD 23288   & $K$  & 0.228  & 0.007 &                  HD 28406   & $K$  & 0.277  & 0.007 \\                   
HD 23324   & $R$  & 0.198  & 0.006 &                  HD 32630   & $H$  & 0.388  & 0.040 \\                   
HD 23338   & $H$  & 0.320  & 0.030 &                  HD 36667   & $H$  & 0.284  & 0.007 \\                   
HD 23441   & $R$  & 0.159  & 0.005 &                  HD 36667   & $K$  & 0.284  & 0.007 \\                   
HD 23753   & $R$  & 0.222  & 0.006 &                  HD 218235  & $H$  & 0.390  & 0.009 \\                   
HD 23753   & $H$  & 0.226  & 0.006 &                  HD 218235  & $K$  & 0.390  & 0.009 
\enddata
\tablecomments{Except where noted, the calibrator uniform disk diameters (UDD)  were adopted from the JMMC Stellar Diameters Catalogue \citep{Bourges:2014}.}
\tablenotetext{a}{We adopted an angular diameter for HD~22860 based on the Gaia parallax of $5.82 \pm 0.11$~mas and an estimated radius of $2.60 \pm 0.13~R_{\odot}$ from \citet{Kervella:2019}.}
\tablenotetext{b}{We adopted the limb-darkened angular diameter of HD~24398 from \citet{Schaefer:2016}, and converted it to a uniform disk diameter in the $H$ band using a limb-darkening coefficient of 0.1792 \citep{Claret:2011}, based on an effective temperature of 21,950 K and $\log g = 3.061$ \citep{Huang:2008}.}
\end{deluxetable}

\subsection{Other Astrometric Observations}
\label{sec:otherastrometry}

Additional long-baseline interferometric observations of Atlas were
reported by \cite{Pan:2004}. They were collected between 1989 and 1992
with the Mark~III stellar interferometer (Mount Wilson, California),
and in 1996--1999 with the Palomar Testbed Interferometer (PTI, Mount
Palomar, California). They were presented in {Cartesian} coordinates
($\Delta\alpha$, $\Delta\delta$) for
the epoch of observation.\footnote{The published signs of both coordinates turn
  out to be reversed. This comes from the fact that the
  interferometric squared visibilities have an inherent
  180\arcdeg\ ambiguity in the position angles.\label{foot:footnote1}}  \cite{Zwahlen:2004}
reported further interferometric observations from the Mark~III
instrument between 1989 and 1992, and additional ones gathered with
the Navy Prototype Optical Interferometer (NPOI, Flagstaff, Arizona)
in 1991--2000. These were provided in polar coordinates ($\rho$,
$\theta$), with their
corresponding error ellipses. We incorporated all of these
measurements into our analysis as published.

Lunar occultation detections of the companion to Atlas have been made
by a number of authors between 1968 and 1991. While such measurements
only yield the projection of the true separation of the binary along
the direction of the Moon's motion ($\rho_{\rm occ}$), they can still provide constraints
that can be used to strengthen the orbit, as we do here.  The
usefulness of these observations was also mentioned by
\cite{Pan:2004}, although they only employed them as a consistency
check on their interferometrically derived orbit. A complete list of
these observations is given in Table~\ref{tab:occ}, to correct several
misprints or mistakes found in the tabulation of \cite{Pan:2004}.

\setlength{\tabcolsep}{4pt}
\begin{deluxetable}{lccccc}
\tablewidth{0pc}
\tablecaption{Lunar Occultation Measurements for Atlas \label{tab:occ}}
\tablehead{
\colhead{JD} &
\colhead{Year} &
\colhead{$\theta_{\rm occ}$} &
\colhead{$\rho_{\rm occ}$} &
\colhead{Phase} &
\colhead{Source}
\\
\colhead{(2,400,000+)} &
\colhead{} &
\colhead{(\arcdeg)} &
\colhead{(mas)} &
\colhead{} &
\colhead{}
}
\startdata
  40221.5734   &  1968.998  &   236*    &   $6.2 \pm 0.4$      &  0.3962  &  1,2  \\
  41314.6075*  &  1971.991  &   128.7   &   $2.2 \pm 0.5$      &  0.1523  &  3    \\
  41396.3724   &  1972.215  &   273.7   &   $2.5 \pm 0.4$      &  0.4333  &  4    \\
  41396.3844   &  1972.215  &   124.0   &   $7.4 \pm 0.4$      &  0.4333  &  4    \\
  41724.7154   &  1973.114  &   228.6*  &   $4.0 \pm 0.4$      &  0.5616  &  2    \\
  47106.5562   &  1987.848  &    21.09  &   $6.8 \pm 0.2$      &  0.0556  &  5    \\
  47106.5562   &  1987.848  &    21.09  &   $6.1 \pm 0.4$      &  0.0556  &  5    \\
  48309.2844   &  1991.141  &   297.45  &   $6.2 \pm 0.3$      &  0.1887  &  5    \\
  48336.5221*  &  1991.216  &   264     &  $13.1 \pm 0.4$\phn  &  0.2823  &  6   
\enddata

\tablecomments{JD epochs marked with an asterisk indicate observations flagged as
  poor by the authors, which we have not used. We note, however,
  that the last one may simply suffer from a misprint: a vector separation
  of $\rho_{\rm occ} = 3.1$~mas, instead of 13.1~mas, would fit our
  final model almost perfectly. Vector angles
  $\theta_{\rm occ}$ marked with an asterisk have been adjusted here
  by 180\arcdeg. All angles are for the equinox of the date of
  observation. Separation vectors $\rho_{\rm occ}$ with no reported
  uncertainties are assigned an initial error of 0.4~mas. Orbital
  phases are based on the ephemeris in Table~\ref{tab:mcmc}. Source
  codes in the last column are:
(1) \cite{Evans:1971};
(2) \cite{McGraw:1974};
(3) \cite{deVegt:1976};
(4) \cite{Bartholdi:1975};
(5) \cite{Meyer:1995};
(6) \cite{Hill:1996}.}

\end{deluxetable}
\setlength{\tabcolsep}{6pt}

The Hipparcos catalog \citep{ESA:1997} reported the first reasonably
accurate orbital solution for Atlas (source identifier HIP~17847),
with a period of $290.7 \pm
8.6$~d.\footnote{An earlier claim of a spectroscopic orbit by
  \cite{Abt:1965} had a period of 1255~d, but proved to be erroneous.}
It was based on the motion of the center of light of
the binary on the plane of the sky, as the system was not spatially
resolved. It also assumed that the orbit is circular,
due to a lack of constraints.  The Hipparcos intermediate astrometric data
(``abscissa residuals'') used for that determination are publicly
available, and have typical
individual uncertainties for Atlas of 1--2~mas that make them useful in
our analysis for constraining the astrometric orbit. We will therefore make
use of these observations below in Section~\ref{sec:orbit}.

We note, finally, that the existence of a wider companion to Atlas
($\rho \leq 0\farcs8$), roughly 3~mag fainter than the combined light
of the close pair, has been claimed since at least 1827
\citep[as reported by][]{Struve:1837}. The Washington Double Star Catalog
\citep[WDS;][]{Worley:1997, Mason:2001} contains several claimed
detections through 1929, and many more instances when it was not
detected.  Many of the most prominent visual binary observers of the
19th and 20th centuries never saw it double. Similar attempts at detection
since the 1970s by the speckle technique, as well as more recently with
adaptive optics, have also failed, even though the companion should have
been a relatively easy target.  Examination of the claimed
detections, kindly provided by R.\ Matson (U.S.\ Naval Observatory),
shows that the position angles and separations scatter considerably,
as indicated also in the notes to the WDS.  Although occasional
references to the existence of this companion, and even the use of its
claimed properties, are still found in the recent literature
\citep[e.g.,][]{Neiner:2015, Gordon:2019}, at the present time we do not regard the
evidence to be sufficiently convincing to support its reality.

\section{Spectral Disentangling}
\label{sec:disentangling}

\begin{figure}
\epsscale{1.15}
\plotone{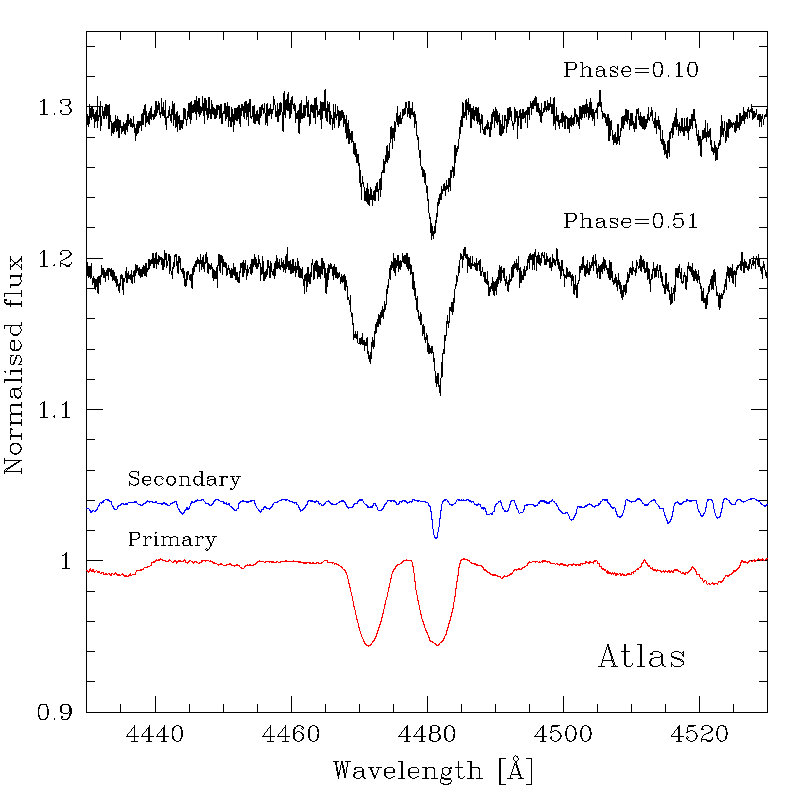}
\figcaption{\emph{Top:} Two of our observed spectra of Atlas (black)
  at opposite quadratures (maximum velocity separation).
  \emph{Bottom:} Disentangled spectra of the binary components in the
  spectral region 4430--4530~\AA, centered on the \ion{He}{1} 4471~\AA\ and \ion{Mg}{2} 4481~\AA\ lines. The secondary component is shifted upward
  by 0.04, for clarity. \label{fig:disent_spec}}

\end{figure}

As is common in double-lined binaries, the spectra of Atlas display a
complex pattern of overlapping lines from the two components that
changes with orbital motion. In Atlas, this is exacerbated by the rapid
rotation of the stars, particularly the primary, such that the lines
of the two objects are never completely separated at any orbital phase.
Two examples at opposite quadratures are shown in black at the top of
Figure~\ref{fig:disent_spec}. The broader features of the primary
star dominate, and have the smaller RV amplitude of the two, while the
secondary can be identified by the superimposed weaker and narrower
lines.

For this work, we made use of the method of spectral disentangling
(SPD) to isolate the individual spectra of the components for further
analysis. In typical uses of SPD \citep{Simon:1994, Hadrava:1995}, the
orbital elements of the binary system and the components' spectra are
calculated simultaneously, in a self-consistent manner. The RVs are
by-passed, and the orbital parameters are optimized directly.  Here,
however, the orbital motion is also strongly constrained by the
astrometric observations described previously. In order to take
advantage of that information, we have therefore chosen to use SPD in
pure separation mode \citep[e.g.,][]{Pavlovski:2010}. In this approach,
the orbital elements are not varied, but are instead held fixed from a
separate orbital analysis that in this case combines the astrometry with RVs,
as we describe later in Section~\ref{sec:orbit}. This avoids the
inconsistency that would arise from the use of a different set of spectroscopic
orbital elements for the reconstructed spectra than those adopted for the rest
of the paper. Here we fixed the orbital elements to those presented below in
Section~\ref{sec:orbit}.

The disentangling code applied in this work, {\sc FDBinary}
\citep{Ilijic:2004}, uses the Fast Fourier Transform and makes
the selection of spectral segments for SPD very flexible, also
allowing one to closely preserve the original resolution of the
observed spectra.  This is particularly important in order to keep as
much wavelength space as possible at the edges of the \'echelle orders
(a prerequisite of SPD in the Fourier domain), which is already
limited due to heavy line blending given the broad spectral features of
the primary.

We applied SPD over the spectral range 4000--5600~\AA, in segments of
50 to 100~\AA. For regions containing the broad Balmer lines, the
segments were 150--200~\AA\ wide. The bottom section of
Figure~\ref{fig:disent_spec} shows a $\sim$100~\AA\ portion of the
disentangled spectra centered on the \ion{He}{1} 4471~\AA\ and
\ion{Mg}{2} 4481~\AA\ lines, which are very prominent in the
primary. A significant difference is seen in the projected rotational
velocities of the two stars, as well as in their relative flux
contributions.

In eclipsing systems in which the flux ratio between the components
changes appreciably over the course of the orbital cycle, it is
possible to infer the proper normalization factors giving the
correct continuum flux of one component relative to the other
\citep[see, e.g.,][]{Pavlovski:2009, Pavlovski:2022, Pavlovski:2023}.
That is not the case for Atlas, however, so there is an inherent
ambiguity in reconstructing the component spectra. They are separated
correctly, i.e., the shapes and relative strengths of the spectral
lines are as they should be for each star, but their fractional light
contribution cannot be determined without external information. We
discuss this further in the following section.

\section{Atmospheric Parameters}
\label{sec:atmospheric}

For the analysis of the disentangled spectra, we employed the method of spectral synthesis as implemented in the {\sc gssp} software package \citep{Tkachenko2015}. This package is designed for the analysis of spectra of single stars and binary systems, and consists of three modules. Two of them were used in this work: {\sc gssp\_single} and {\sc gssp\_binary}. The former allows for the analysis of the disentangled spectra of double-lined binaries, but treats each component as if
it were an isolated single star, and treats light dilution in the disentangled spectra as a wavelength-independent effect. This implies that the parallel analysis of the binary components' disentangled spectra may result in a total light factor that either exceeds or falls short of unity. The {\sc gssp\_binary} module, on the other hand, couples the components’ spectra and computes the light contribution of each star per wavelength bin. In this case, the light factors are replaced by a single free parameter — the squared radius ratio between the two stars — as shown by \citet[][his Eqs.\ 3 and 4]{Tkachenko2015}.

Because the {\sc gssp\_single} module is generally faster, it is a good choice for obtaining initial estimates of the atmospheric properties of the binary components, albeit under the assumption of a wavelength-independent light dilution factor. Once adequate initial guesses are determined for both components, the more computationally demanding but physically more accurate {\sc gssp\_binary} module can be used. The {\sc gssp} package employs the {\sc SynthV} radiative transfer code \citep{Tsymbal1996} to compute synthetic spectra over arbitrary wavelength ranges and for arbitrary surface compositions. To generate theoretical spectra, {\sc SynthV} relies on atmosphere models — in our case, a pre-computed grid of {\sc LLmodels} \citep{Shulyak2004} as published by \citet{Tkachenko2012}.

Our initial analysis of the disentangled spectra using the {\sc gssp\_single} module revealed the following key features: {\it (i)} although the two stars have similar effective temperatures, their $\log g$ values differ significantly, indicating that the primary component is more evolved; {\it (ii)} the primary displays substantially broader spectral lines than the secondary, suggesting a large difference in rotational velocities — unless the secondary is seen at a low inclination relative to the line of sight; {\it (iii)} a striking mismatch in metallicity is found between the components, with the secondary exhibiting ${\rm [M/H]} \approx +0.7$~dex, while the primary shows a nearly solar composition.

The latter result is unexpected for stars born in the same binary system from a common natal cloud, prompting a more detailed investigation. Upon closer inspection of the spectral fits, we found that the high metallicity estimate for the secondary was largely driven by a systematic increase in the depths of iron, chromium, and titanium lines. Additionally, the secondary’s helium lines appeared significantly weaker than expected for a star with its effective temperature. This is
consistent with an earlier classification of Atlas as a He-weak object
\citep{Renson:2009}.
We therefore fixed the global metallicity of the secondary to the solar value, consistent with the primary, and re-optimized its atmospheric parameters alongside the individual abundances of He, Fe, Cr, and Ti. The resulting atmospheric parameters and elemental abundances from this analysis are listed in Table~\ref{tab:spectro} (columns 2 and 3).

With the above results in hand, we employed the {\sc gssp\_binary} module to analyze the disentangled spectra of the two components simultaneously, replacing the
wavelength-independent light factors with the radius ratio between the stars.
It is well known that strong abundance anomalies for helium and/or metals, such as those found in the secondary, lead to non-negligible changes in the atmospheric structure — which can, in turn, affect the determination of atmospheric parameters, if ignored \citep[e.g.,][]{Lehmann2007,Shulyak2009}. To avoid this bias, 
we employed a custom grid of model atmospheres for the secondary, computed by fixing the abundances of He, Fe, Cr, and Ti to the values obtained from the unconstrained mode analysis.

\begin{figure*}
\includegraphics[width=\textwidth,trim=0.8in 5.6in 0.7in 0.8in]{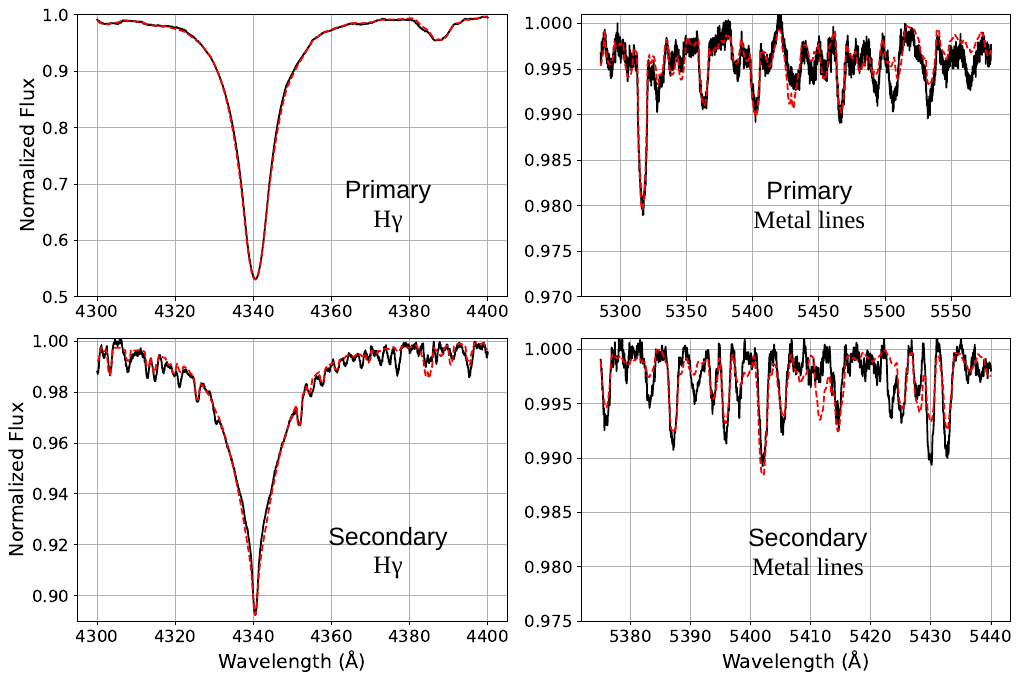}
\figcaption{Comparison between the disentangled spectra (solid black line) and best-fit model (dashed red line) for the primary and secondary components of the Atlas system. Two different wavelength regions are shown, featuring H$\gamma$ and a section rich in metal lines. Note the changing vertical scales in the four panels.
\label{fig:bestfitspectra}}
\end{figure*}

The results are presented in Table~\ref{tab:spectro} (columns 4 and 5), and are considered the final set of parameters in this study. While both stars are now found to be slightly hotter than suggested by the previous analysis, the elemental abundances of He, Fe, Cr, and Ti in the atmosphere of the secondary remained largely unchanged. Figure~\ref{fig:bestfitspectra} shows a comparison between the disentangled and best-fit model spectra for both components (top row for the primary, bottom row for the secondary), focusing on the H${\gamma}$ line profile (left column) and a set of metal lines (right column).

Our findings leave no doubt that the secondary component of Atlas is a chemically peculiar star. Numerous spectroscopic studies of such objects in the literature report a wide range of surface chemical abundance distributions. Those distributions vary from simple to complex and, in some cases, they appear to correlate with magnetic field topologies \citep[e.g.,][]{Kochukhov2014,Kochukhov2017,Alecian2015}. Observationally, such distributions manifest as temporal variability in the line profiles of the affected elements. In fact, upon close inspection of the average spectral lines of the secondary, we detected changes in the shapes of line profiles with time that are unrelated to the orbital motion of the star. We describe and illustrate
this in more detail in the next section. On the other hand, the method of spectral disentangling employed in Section~\ref{sec:disentangling} assumes the absence of variability in the observed composite spectra, beyond that caused by orbital motion. As a result, the 1$\sigma$ uncertainties listed for the atmospheric parameters of both stars in Table~\ref{tab:spectro} should be regarded as purely statistical errors.

\setlength{\tabcolsep}{11pt}
\begin{deluxetable*}{lcccc}
\tablewidth{0pc}
\tablecaption{Spectroscopic Properties of Atlas \label{tab:spectro}}
\tablehead{
\colhead{} &
\multicolumn{2}{c}{Unconstrained Mode} &
\multicolumn{2}{c}{Constrained Mode} \\
\colhead{Parameter} &
\colhead{Primary} &
\colhead{Secondary} &
\colhead{Primary} &
\colhead{Secondary}
}
\startdata
$T_{\rm eff}$ (K)         &  $12100 \pm 200$\phn\phn  & $12150 \pm 450$\phn\phn   & $12525 \pm 200$\phn\phn  & $12835 \pm 450$\phn\phn \\
$\log g$ (dex)            &  $3.40 \pm 0.05$          & $4.15 \pm 0.11$           & $3.38 \pm 0.07$          & $4.20 \pm 0.15$ \\
$\xi$ (\kms)              &  $0.0^{+1.2}$             & $1.0 \pm 1.5$             & $0.5^{+0.9}_{-0.5}$      & $1.0^{+1.5}_{-1.0}$   \\
$v \sin i$ (\kms)         &  $212 \pm 10$\phn         & $48 \pm 5$\phn            & $217 \pm 9$\phn\phn      & $47 \pm 7$\phn \\
${\rm [M/H]}$             &  $-0.02 \pm 0.06$\phs     & 0.0 (fixed)               & $-0.03 \pm 0.06$\phs     & 0.0 (fixed)    \\
Light factor              & $0.83 \pm 0.03$           & $0.15 \pm 0.03$           & \nodata                  & \nodata \\
$R_1/R_2$ & \multicolumn{2}{c}{\nodata} & \multicolumn{2}{c}{$2.33 \pm 0.07$} \\
${\rm [He}/N_{\rm tot}]$  &  \nodata                  & $-1.98 \pm 0.07$\phs      & \nodata                  & $-2.00 \pm 0.09$\phs \\
${\rm [Fe}/N_{\rm tot}]$  &  \nodata                  & $-3.53 \pm 0.08$\phs      & \nodata                  & $-3.53 \pm 0.08$\phs \\
${\rm [Ti}/N_{\rm tot}]$  &  \nodata                  & $-6.25 \pm 0.40$\phs      & \nodata                  & $-6.35 \pm 0.40$\phs \\
${\rm [Cr}/N_{\rm tot}]$  &  \nodata                  & $-5.43 \pm 0.25$\phs      & \nodata                  & $-5.45 \pm 0.40$\phs 
\enddata
\tablecomments{$\xi$ represents the microturbulent velocity. The element abundances in the constrained mode above, expressed relative to the Sun, are as follows:
He ($-0.90$~dex), Fe (+1.05~dex), Ti (+0.80~dex), and Cr (+0.95~dex).
Results from the constrained mode are adopted for the remainder of this work.}
\end{deluxetable*}
\setlength{\tabcolsep}{6pt}

\begin{figure}
\epsscale{1.17}
\plotone{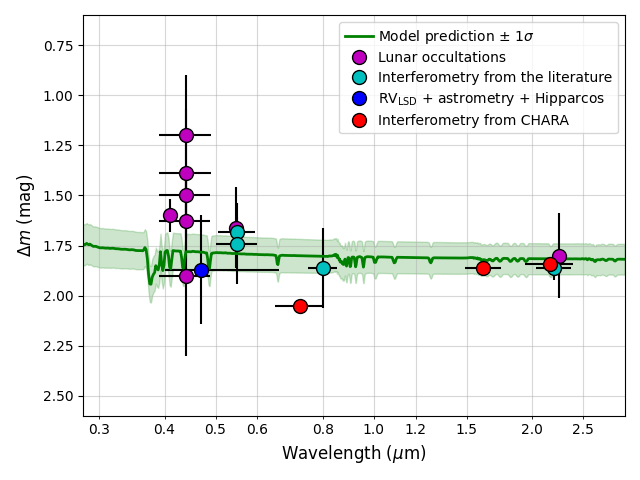}
\figcaption{Measured magnitude differences between the components
of Atlas, together with the prediction from our spectroscopic analysis in
Section~\ref{sec:atmospheric}. The shaded area around the curve is the
1$\sigma$ uncertainty range obtained by propagating the errors in
the stellar temperatures, surface gravities, and radius ratio.
The horizontal error bars on the measurements represent the
width of the bandpass.\label{fig:dmag}}
\end{figure}

A sanity check on the spectroscopic parameters for Atlas may be obtained
by comparing the predicted magnitude difference $\Delta m$ between the
components against independent measurements from astrometry. To
predict $\Delta m$ as a function of wavelength, we used synthetic
spectra based on Kurucz model atmospheres \citep{Castelli:2003}
computed for solar metallicity and the components' derived properties
($T_{\rm eff}$, $\log g$), scaling the fluxes by the derived radius
ratio.  Empirical measurements of the magnitude difference are
available from the lunar occultation observations of Atlas, the
interferometric observations by \cite{Pan:2004}, our CHARA
observations reported in Section~\ref{sec:orbit} at three wavelengths
from the PAVO, CLIMB, MIRC-X, and MYSTIC beam combiners, and a value
in the Hipparcos bandpass, also reported later.
The model predictions are compared against these measurements
in Figure~\ref{fig:dmag}. The agreement is generally very good, with
the exception of a few of the early lunar occultation measurements,
which have large formal uncertainties.  The two that deviate the most
come from observations that were reported as problematic in the
original publications (poor seeing, etc.). We view the good agreement
as broadly supportive of the accuracy of our spectroscopic determinations.

\section{Radial Velocities}
\label{sec:rvs}

Inferring {RVs} with high precision benefits from data with high signal-to-noise ratios (S/N). Since all spectral lines in binary star spectra are affected by the Doppler effect in the same way, it is advantageous to apply spectral line averaging techniques to maximally enhance the S/N of the resulting mean profile used for the RV analysis. In this work, we employed the method of Least-Squares Deconvolution \citep[LSD,][]{Donati1997}, as implemented by \citet{Tkachenko2013}. In practice, the LSD method performs a ``deconvolution'' of all spectral lines that are common between the observed spectrum of the target and a pre-computed line list, and then computes a mean profile from all these “deconvolved” spectral features. There are two key limitations to consider when pre-computing the line list:
{\it (i)} all spectral lines should have similar shapes, which means that hydrogen and helium lines should be excluded due to their pressure-dominated broadening mechanisms; {\it (ii)} the weakest lines are also best excluded to avoid introducing significant noise into the resulting mean profile.

Given that the components of the Atlas system have similar spectral types, and that the primary contributes a larger share of the total light, we prepared a single line list by extracting relevant information from the Vienna Atomic Line Database \citep[VALD,][]{Piskunov1995,Kupka1999}, using the atmospheric parameters of the primary derived in Section~\ref{sec:atmospheric} and listed in Table~\ref{tab:spectro} (column~4). In our selection, we discarded all spectral lines with predicted line strengths weaker than 0.03 (i.e., 3\% depth in continuum units). This threshold is dictated by the high rotational velocity of the primary component, which causes weaker lines to be buried in the noise, contributing no useful signal to the calculation of the LSD profiles.

\begin{figure}
\epsscale{1.17}
\plotone{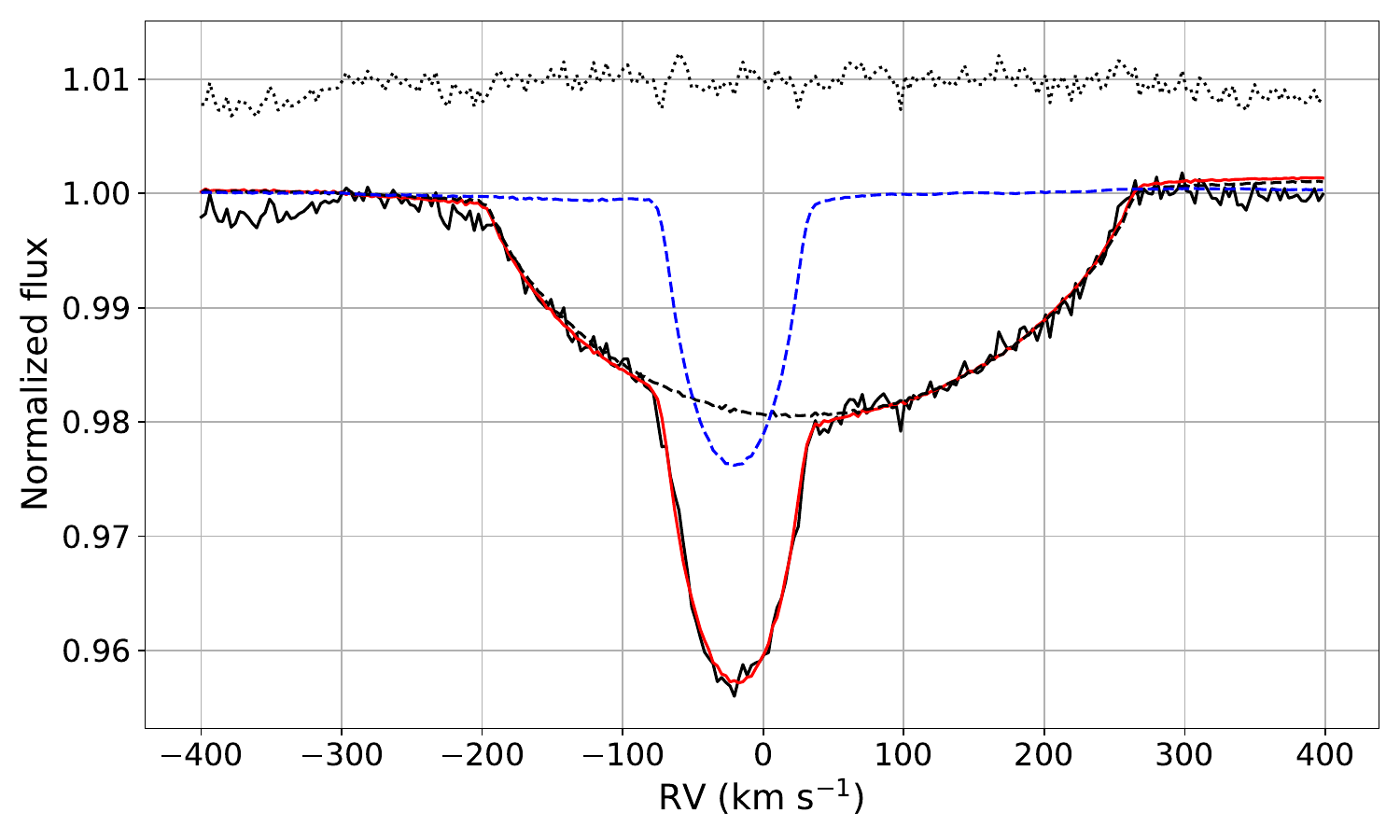}
\figcaption{An observed composite LSD profile (black solid line) fitted with a superposition of two model LSD profiles (red solid line). The black and blue dashed lines show the respective model LSD profiles of the primary and secondary components. Residuals, obtained by subtracting the composite model profile from the observations, are shown with a black dotted line at the top, shifted upward by 0.01, for clarity.
\label{fig:lsdfit}}
\end{figure}

Each LSD profile from the time series, computed in this way, was analyzed to infer the RVs of both components. A separate grid of synthetic LSD profiles was computed
for each star based on the values of $T_{\rm eff}$, $\log g$, [M/H], and
$\xi$ as listed in Table~\ref{tab:spectro}. The grids span a range of projected rotational velocity of $v \sin i = {\rm [180, 260, 2]}~\kms$ for the primary and $v \sin i = {\rm [25, 75, 1]}~\kms$ for the secondary. Each observed LSD profile was then fitted with a superposition of two synthetic LSD profiles — one for each binary component — with a total of six adjustable parameters: their positions (represented by the RVs), widths (represented by $v \sin i$), and depths (represented by line-depth scaling factors). 
Our fitting procedure is conceptually similar to fitting a superposition of two Gaussians with variable positions, widths, and amplitudes, but different in that our model is based on detailed radiative transfer computations rather than an arbitrary analytical function. An example of a fit to one of the LSD profiles is shown in Figure~\ref{fig:lsdfit}. RVs derived in this way for both binary components are listed in Table~\ref{tab:rvs}.

\setlength{\tabcolsep}{4pt}
\begin{deluxetable}{lcccc}
\tablewidth{0pc}
\tablecaption{CfA Radial Velocities of Atlas \label{tab:rvs}}
\tablehead{
\colhead{BJD} &
\colhead{$RV_1$} &
\colhead{$RV_2$} &
\colhead{S/N} &
\colhead{Phase}
\\
\colhead{(2,400,000+)} &
\colhead{(\kms)} &
\colhead{(\kms)} &
\colhead{} &
\colhead{}
}
\startdata
    58050.8325  &   $33.4 \pm 3.9$\phn      &  $-20.6 \pm 1.6$\phn\phs  &   686  &   0.6531 \\
    58079.8815  &   $23.8 \pm 3.0$\phn      &  $-12.5 \pm 1.7$\phn\phs  &   484  &   0.7529 \\
    58109.8188  &   $10.6 \pm 4.6$\phn      &    $7.1 \pm 1.4$          &   542  &   0.8558 \\
    58127.5616  &   $-3.6 \pm 3.5$\phs      &   $22.7 \pm 1.7$\phn      &   641  &   0.9168 \\
    58142.6446  &  $-16.4 \pm 5.1$\phn\phs  &   $40.5 \pm 1.7$\phn      &   463  &   0.9686 
\enddata

\tablecomments{S/N represents the signal-to-noise ratio per resolution element.
Orbital phases are based on the ephemeris of
  Table~\ref{tab:mcmc}. (This table is available in its entirety in
  machine-readable form.)}

\end{deluxetable}

\begin{figure}
\epsscale{1.17}
\plotone{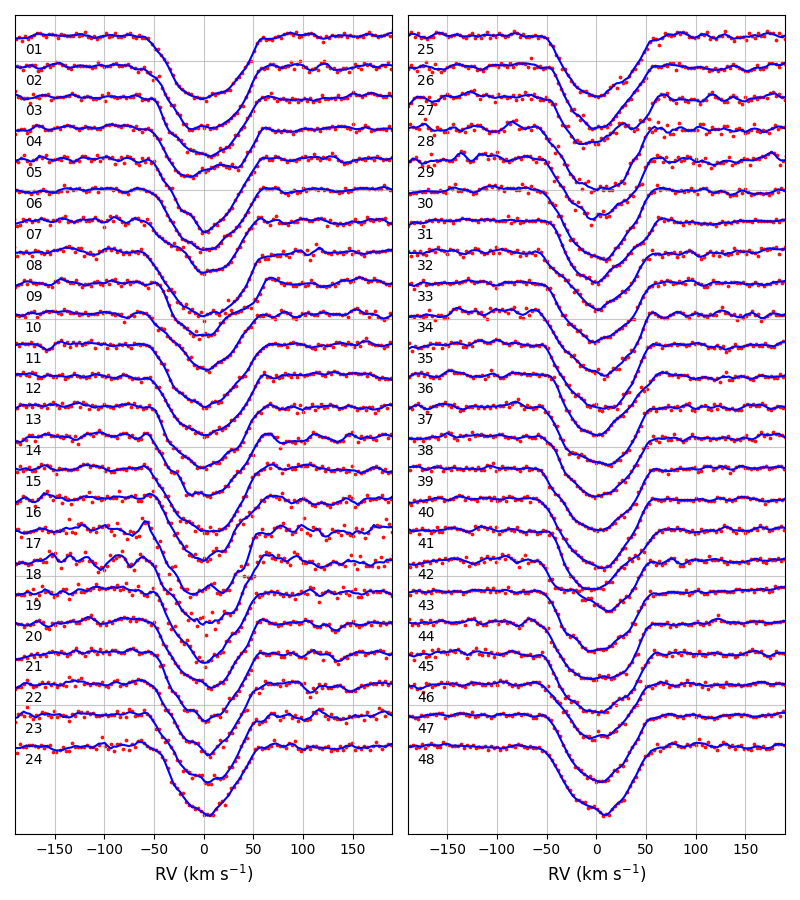}
\figcaption{LSD profiles of the secondary component, shifted to zero velocity, obtained after subtracting the primary's contribution from the composite average profiles. The red dots are the measurements, and the solid blue lines represent spline fits to the profiles. The spectra are numbered by date, in the same order as they appear in Table~\ref{tab:rvs}.\label{fig:lsdsec}}
\end{figure}

As mentioned in the previous section, the anomalous abundances of He, Fe, Cr, and Ti found in the secondary of Atlas manifest themselves in the spectrum through distortions in the line profiles that change with time. To visualize these distortions, we subtracted the best-fit average profiles of the primary — obtained in the previous step — from the observed composite LSD profiles of the system at each epoch. The resulting average profiles of the secondary are shown in Figure~\ref{fig:lsdsec}. The changes in shape are clearly visible to the eye, and are suggestive of rotational modulation caused by surface inhomogeneities on the secondary star. Inverting these variations to infer detailed surface abundance maps is beyond the scope of this study, and would be challenging with the sparse
sampling we have. Nevertheless, it is reasonable to expect that the distortions will increase the scatter in the velocities of the secondary (see the next section), and it is therefore likely that the formal RV uncertainties for that star reported in Table~\ref{tab:rvs} are somewhat underestimated.

\section{Orbital Analysis}
\label{sec:orbit}

\subsection{Procedure and Treatment of Errors}

A joint orbital analysis of the {RVs} and the astrometry
for Atlas was carried out within a Markov chain Monte Carlo framework,
using the {\sc emcee}
package\footnote{\url{https://emcee.readthedocs.io/en/stable/index.html}}
of \cite{Foreman-Mackey:2013}. For consistency, all position angles
were precessed to the year J2000.0, where needed. The orbital elements describing the
relative motion on the plane of the sky are the period ($P$), the
angular semimajor axis ($a^{\prime\prime}$), the eccentricity ($e$)
and argument of periastron for the secondary ($\omega_2$), the orbital
inclination angle ($i_{\rm orb}$), the position angle of the
ascending node for J2000.0 ($\Omega$), and a reference time of
periastron passage ($T_{\rm peri}$). For numerical efficiency, we expressed
the eccentricity and argument of periastron as $\sqrt{e}\cos\omega_2$ and
$\sqrt{e}\sin\omega_2$, and the inclination angle as $\cos i_{\rm orb}$.
The three spectroscopic orbital
elements are $K_1$, $K_2$, and $\gamma$, which represent the velocity
semiamplitudes and center-of-mass velocity. 

Although the Hipparcos mission did not spatially resolve the Atlas binary, it
did detect the astrometric wobble due to the companion,
and was able to follow the path of the center of light of the system,
with the simplifying assumption that the orbit is circular. That path
{(once proper motion and the parallactic motion are removed)}
is simply a scaled-down version of the relative orbit, reflected around
the center of mass, but otherwise with the same shape and orientation.
Here we have incorporated the {one-dimensional} intermediate astrometric data from the satellite
(abscissa residuals) directly into our modeling of the system, which
allows us to remove the assumption of circularity, but
introduces several additional free parameters. One
is the angular semimajor axis of the photocenter
($a^{\prime\prime}_{\rm phot}$). Another five new parameters are corrections to the catalog
values of the position and proper motion components, and parallax ($\Delta\alpha^*$,
$\Delta\delta$, $\Delta\mu_{\alpha}^*$, $\Delta\mu_{\delta}$,
$\Delta\pi_{\rm Hip}$), where the notation $\Delta\alpha^* =
\Delta\alpha \cos\delta$ and $\Delta\mu_{\alpha}^* = \Delta\mu_{\alpha}
\cos\delta$. For details of the formalism for
incorporating these observations into an orbital solution, see, e.g.,
\cite{Torres:2007}, and references therein. 

The observations for Atlas included in
our analysis are the 19 pairs of ($\Delta\alpha^*$, $\Delta\delta$)
observations from \cite{Pan:2004}, the 12 pairs of ($\rho$, $\theta$)
measurements by \cite{Zwahlen:2004}, 7 lunar occultation measurements,
our 48 pairs of RVs, and 36 Hipparcos abscissa residuals.
Not all of the interferometric
measurements from CHARA allow for the determination of relative
positions on a nightly basis, due to insufficient $uv$ coverage in
some cases. For this reason, we incorporated all the (squared)
visibilities from PAVO, CLIMB, MIRC-X, and MYSTIC directly into our
analysis, along with closure phases where available. For MIRC-X and
MYSTIC, the recorded wavelengths were adjusted by dividing by the recommended
correction factors $1.0054 \pm 0.0006$ and $1.0067 \pm 0.0007$, respectively, following \cite{Gardner:2022} and Gardner\,(2022, priv.\ comm.).
Bandwidth smearing was accounted for following \cite{Gardner:2021}.

The use of these data requires us to solve for the flux ratio $F_2/F_1$ in the
three different CHARA wavelength bands (PAVO, $H$, and $K$), as well as for
the angular diameter of the primary ($\phi_1$), which is
resolved in these observations. \cite{Gordon:2019} measured
$\phi_1 = 0.464 \pm 0.043$~mas using CLIMB. The smaller secondary is unresolved, so we
held its angular diameter fixed at a value of $\phi_2 = 0.20$~mas,
{based on its estimated radius and the distance (see below)}. To
reduce the number of free parameters, the primary diameter that we solved
for is a limb-darkened value ($\phi_{\rm 1,LD}$) common to all three
bandpasses, using the appropriate visibility function for a linear
limb-darkened stellar disk model from \cite{HanburyBrown:1974}.
Limb-darkening coefficients were taken from the tabulation
of \cite{Claret:2011}, for the properties of the stars as determined
earlier.\footnote{The limb-darkening coefficients in the PAVO, $H$, and $K$ bands are
0.312, 0.164, and 0.138 respectively for the primary, and 0.295, 0.149, and 0.126 for the
secondary.}

The line profile distortions that are obvious in the secondary of
Atlas can potentially introduce a bias in the velocities, such that the RV
zeropoints for the primary and secondary may not be the same. To
prevent this from affecting the solution, we introduced an additional free parameter
($\Delta_{\rm RV}$) to represent a possible difference between the zeropoints.
Additionally, as the formal errors for the observations may not always
be accurate, and to ensure proper weighting of the different datasets,
we solved for additional parameters in the form of multiplicative
scale factors for the uncertainties, separately for each type of
observation: two for the {RVs} ($f_{\rm RV1}$, $f_{\rm RV2}$),
two others ($f_{\rm P}$, $f_{\rm Z}$) for the interferometric measurements of
\cite{Pan:2004} and \cite{Zwahlen:2004}, respectively, one more for
the lunar occultation measurements ($f_{\rm occ}$), and another for
the Hipparcos abscissa residuals ($f_{\rm Hip}$).
Visibility calibration uncertainties from the MIRC-X
and MYSTIC beam combiners have been estimated to be at the level of
about 5\%, and the formal $V^2$ errors already incorporate this
contribution. For PAVO, they are estimated to be roughly 3\% (J.\ Jones,
priv.\ comm.), but are not factored into the formal errors.
Initial checks for PAVO and CLIMB suggested uncertainties
somewhat larger than those produced by the reduction pipeline for these instruments.
To account for this, we solved for a contribution $\sigma_{V^2}$ added in
quadrature to the internal errors. For the closure phases, we solved
for similar jitter terms $\sigma_{\rm CP}$ for CLIMB, MIRC-X, and
MYSTIC. In all, we solved for 34 adjustable parameters.

Our MCMC analysis used 100 random walkers with 20,000 links each,
after burn-in. We checked for convergence by examining the chains
visually, and by requiring a Gelman-Rubin statistic of 1.05 or smaller
\citep{Gelman:1992}. The priors for all parameters were uniform over
suitable ranges, except for those of the error scaling factors, which
were log-uniform.

We point out that we did not
enforce equality in our analysis between the final Hipparcos parallax
(i.e., the catalog value plus $\Delta\pi_{\rm Hip}$) and the orbital
parallax, because of the known systematic error in the Hipparcos determinations
for the Pleiades (Section~\ref{sec:introduction}). By allowing the
Hipparcos parallax to be different, the goal was to avoid biasing the
constraint that the abscissa residuals contribute to the orbital motion,
which is our main reason for incorporating those measurements.

\subsection{Model for the Primary Disk}
\label{sec:primarydisk}

After an initial fit as described above, additional evidence prompted us to consider a model for the brightness distribution of the primary
star on the plane of the sky that is different from a circular model assumed initially.
There were two motivations for this.
The first was that earlier
fits to some of the CHARA data alone, intended to solve for the
relative position, flux ratio, and primary angular diameter
on nights with sufficient $uv$ coverage (MIRC-X, MYSTIC), displayed a
larger scatter in the $\phi_1$ values than we expected.  Secondly, our
spectroscopically derived $v \sin i$ value of 217~\kms\ is large
enough that one might expect the star to be rotationally distorted, to
a degree that may even be detectable with our interferometric
measurements. This has previously been achieved in other similarly rapidly
rotating stars such as $\alpha$~Aql (Altair), $\alpha$~Oph, $\alpha$~Cep,
and others \citep[see, e.g.,][and references therein]{vanBelle:2012}.
The effect would provide an explanation for the scatter mentioned
above, as cross-sections of the apparent disk at different baseline
orientations would lead to slightly different values for $\phi_1$.

\begin{figure}
\epsscale{1.17}
\plotone{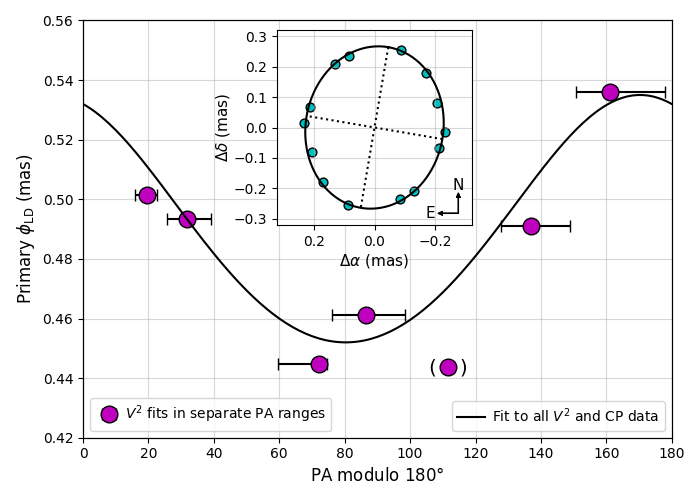}
\figcaption{Limb-darkened diameter of the primary of Atlas as a
function of the baseline orientation (PA). The points represent
independent diameter estimates over separate PA intervals, with the
horizontal error bars indicating the range in each set. 
The measurement in parentheses was based on a small number of
visibilities, and is very uncertain. Formal
diameter error bars are typically smaller than the point size, and are not
shown. The curve is not a fit to these measurements, but is instead a
fit to all CHARA visibilities and closure phases with an elliptical
model for the stellar disk (see the text). The inset shows the
projection of the stellar shape on the plane of the sky, along with
the measurements (duplicated, given that interferometric visibilities
are invariant under a 180\arcdeg\ rotation around the center).
\label{fig:diam}}
\end{figure}

In addition to having the highest spatial resolution among the CHARA
measurements of Atlas, the PAVO data were taken over different and
relatively small ranges of baseline orientations on each night. 
While this limited $uv$ coverage prevents us from measuring nightly
relative positions,
it makes these data ideally suited for exploring possible differences in
$\phi_1$. An MCMC solution was carried out in which we used
all other non-CHARA observations in order to constrain the orbit, and
solved for a separate primary diameter from the PAVO measurements on each of the eight
nights in which that instrument was used. A suggestive trend was seen, with the diameters reaching a
minimum near the middle of the range of position angles
(i.e., baseline orientations). To improve
the sampling, we then merged together the visibilities from all of the
beam combiners, split them into seven 10--30\arcdeg\ intervals
of baseline orientation, and repeated the exercise.
The resulting limb-darkened diameters as a
function of the mean position angle of the baseline are shown in
Figure~\ref{fig:diam}. They support the results from PAVO alone, and
lead us to conclude that the non-circular shape of the primary of Atlas is
indeed detectable.

To first order, a star distorted by rapid rotation will have a
projected shape on the plane of the sky that is similar to an
ellipse. We therefore changed our initial model of a circular
limb-darkened disk for the primary to one featuring a limb-darkened elliptical
disk. For convenience, we used the polar form of the equation for
an ellipse to express the diameter measured along a
given orientation as
\begin{equation}
\label{eq:1}
    \phi_1({\rm PA}) = \frac{\phi_{\rm min}}{\sqrt{1 - e^2_{\phi} \cos^2(\rm{PA}-\theta_{\phi})}}~,
\end{equation}
where $\phi_{\rm min}$ represents the minor axis of the ellipse,
$\theta_{\phi}$ is the orientation angle of the major axis relative to
the north, and PA is the position angle of the projected baseline
(east of north). The eccentricity of the ellipse, $e_{\phi}$, has the
expression $e_{\phi} = \sqrt{1 - (\phi_{\rm min}/\phi_{\rm maj})^2}$
in terms of the major and minor axes.

For our final MCMC solution, we solved for the three parameters of the
apparent elliptical disk together with the rest of the elements
mentioned earlier, incorporating all CHARA visibilities and closure
phases directly. We obtained $\phi_{\rm min} = 0.4523 \pm 0.0026$~mas,
$e_{\phi} = 0.534 \pm 0.012$, and $\theta_{\phi} = -9\fdg9 \pm
1\fdg5$. The corresponding axial ratio is $\phi_{\rm min}/\phi_{\rm
maj} = 0.8456 \pm 0.0076$.
An illustration of the shape and size of the
ellipse on the plane of the sky is shown in the inset in Figure~\ref{fig:diam}.

The above model is only an approximation to the true stellar shape, and does
not account for effects such as gravity darkening. This 
may not be negligible for Atlas, and it introduces an asymmetry.
Because of this, the formal errors reported above may not reflect
the true uncertainties of the primary's disk parameters.
Nevertheless,
the results serve to show for the first time that the primary in the Atlas
system has a detectable oblateness caused by its rapid rotation. More
realistic models where all physical effects are properly considered
are certainly possible, and have been applied to other fast
rotators \citep[see, e.g.,][]{vanBelle:2001, Domiciano:2003,
Aufdenberg:2006, Monnier:2007, Zhao:2009}. Such an analysis for Atlas
is beyond the scope of the present work, and would
benefit from additional interferometric observations.

\subsection{Results}

\begin{figure}
\epsscale{1.17}
\plotone{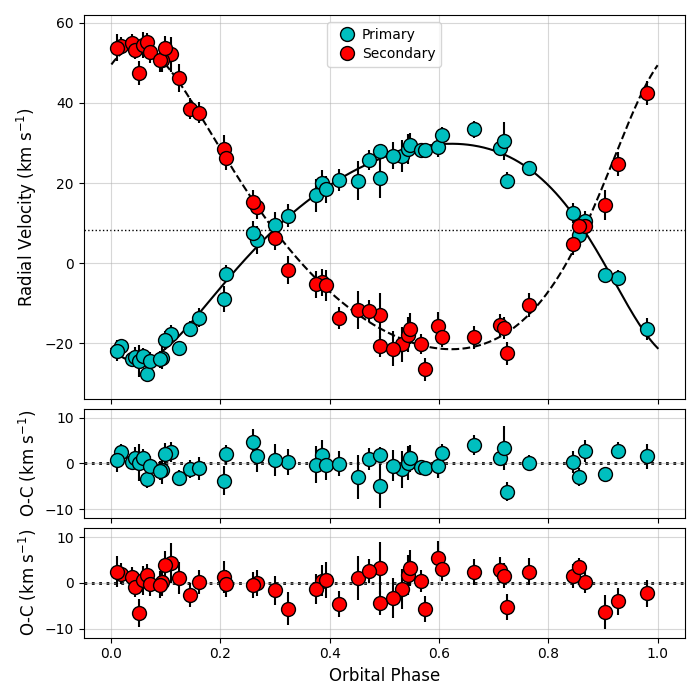}
\figcaption{{RV} measurements for Atlas with our model. The
  center-of-mass velocity is indicated with the dotted line. Residuals
  are shown at the bottom. \label{fig:rvorbit}}
\end{figure}

\begin{figure}
\epsscale{1.17}
\plotone{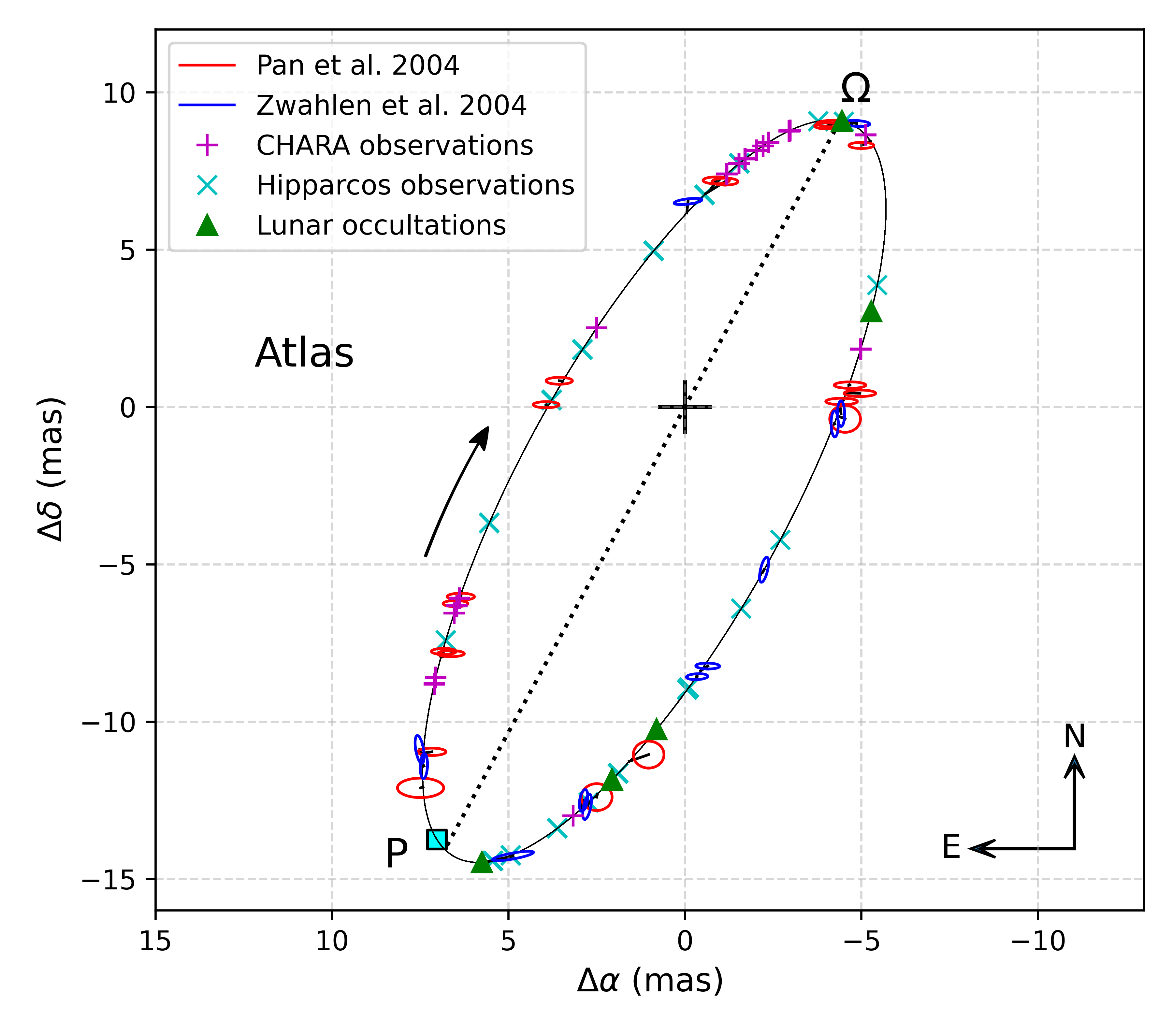}
\figcaption{Interferometric observations of Atlas from \cite{Pan:2004}
  (shown in red) and \cite{Zwahlen:2004} (blue). Error ellipses represent
  the uncertainty on each axis, and short line segments connect the
  measurement with the predicted position on the orbit. The
  one-dimensional lunar occultation measurements cannot be shown here,
  and are represented with triangles at their predicted location from the
  model, merely to illustrate their phase coverage. Similarly with the CHARA
  visibilities and closure phases, and with the one-dimensional Hipparcos
  observations. The predicted locations of the Hipparcos measurements
  are shown here on the relative orbit rather than on the scaled-down
  photocenter orbit, as the phase coverage is of course the same.
  Collectively, the phase coverage of the astrometric observations
  is near complete.
  The dotted line indicates the line of nodes, with the
  ascending node marked as ``$\Omega$". The square labeled ``P''
  indicates periastron, coincidentally very near the descending
  node. \label{fig:visualorbit}}
\end{figure}

Our orbital elements for Atlas from our final MCMC solution that uses
the elliptical model for the shape of the primary
are presented in Table~\ref{tab:mcmc}, along with derived
properties for the Atlas system, including the component masses and the
orbital parallax.\footnote{While our value for $\Omega$ follows the
  usual convention, and represents the node where the secondary is
  receding from the observer, the angle reported by \cite{Pan:2004} is
  flipped by 180\arcdeg\ (see footnote~\ref{foot:footnote1}). The
  angles $\omega$ and $\Omega$ by \cite{Zwahlen:2004} both need
  to be changed by 180\arcdeg\ to be in the right quadrant.}
The {RVs} and our model for the spectroscopic orbit are
shown in Figure~\ref{fig:rvorbit}, and a representation of the
astrometric observations and visual orbit model may be seen in
Figure~\ref{fig:visualorbit}.

As anticipated, we find a systematic velocity {zeropoint} offset of $\Delta_{\rm RV} = 2.15 \pm 0.53~\kms$ between the primary and secondary, which may be related to
the peculiarities in the secondary. We also find that this star's RV uncertainties
require an inflation factor of $f_{\rm RV2} \approx 1.7$ in order to achieve a
reduced $\chi^2$ value near unity, which is likely caused by the line profile variations illustrated previously. On the other hand, the formal RV errors for
the primary reported in Table~\ref{tab:rvs} appear to be overestimated by
about a factor of two.

\setlength{\tabcolsep}{3pt}
\begin{deluxetable}{lcc}
\tablewidth{0pc}
\tablecaption{Orbital Parameters for Atlas \label{tab:mcmc}}
\tablehead{
\colhead{Parameter} &
\colhead{Value} &
\colhead{Prior}
}
\startdata
 $P$ (day)                               & $290.9919 \pm 0.0028$\phn\phn & [250, 300]     \\
 $a^{\prime\prime}$ (mas)                & $12.9896 \pm 0.0036$\phn      & [5, 20]        \\
 $\sqrt{e}\cos\omega_2$                  & $+0.44243 \pm 0.00032$\phs    & [$-1$, 1]      \\
 $\sqrt{e}\sin\omega_2$                  & $-0.19974 \pm 0.00063$\phs    & [$-1$, 1]      \\
 $\cos i_{\rm orb}$                      & $-0.30675 \pm 0.00054$\phs    & [$-1$, 1]      \\
 $\Omega$ (deg)                          & $334.202 \pm 0.025$\phn\phn   & [0, 360]       \\
 $T_{\rm peri}$ (BJD)                    & $50585.988 \pm 0.096$\phm{2222} & [50550, 50600] \\
 $K_1$ (\kms)                            & $27.09 \pm 0.40$\phn          & [15, 50]       \\
 $K_2$ (\kms)                            & $37.63 \pm 0.53$\phn          & [15, 50]       \\
 $\gamma$ (\kms)                         & $+8.41 \pm 0.32$\phs          & [0, 15]        \\
 $\Delta_{\rm RV}$ (\kms)                & $+2.15 \pm 0.53$\phs          & [$-20$, 20]    \\
 $a^{\prime\prime}_{\rm phot}$ (mas)     & $3.43 \pm 0.40$               & [0, 20]        \\
 $\Delta\alpha^*$ (mas)                  & $+0.03 \pm 0.41$\phs          & [$-50$, 50]    \\
 $\Delta\delta$ (mas)                    & $-0.85 \pm 0.28$\phs          & [$-50$, 50]    \\
 $\Delta\mu_{\alpha}^*$ (mas yr$^{-1}$)  & $+0.44 \pm 0.44$\phs          & [$-50$, 50]    \\
 $\Delta\mu_{\delta}$ (mas yr$^{-1}$)    & $+0.07 \pm 0.35$\phs          & [$-50$, 50]    \\
 $\Delta\pi_{\rm Hip}$ (mas)             & $-0.49 \pm 0.46$\phs          & [$-50$, 50]    \\
 $(F_2/F_1)_{\rm PAVO}$                  & $0.1509 \pm 0.0051$           & [0.01, 1]      \\
 $(F_2/F_1)_H$                           & $0.17984 \pm 0.00064$         & [0.01, 1]      \\
 $(F_2/F_1)_K$                           & $0.18308 \pm 0.00054$         & [0.01, 1]      \\
 $\phi_{\rm min}$ (mas)                  & $0.4523 \pm 0.0026$           & [0.1, 1.5]     \\
 $e_{\phi}$                              & $0.534 \pm 0.012$             & [0, 1]         \\
 $\theta_{\phi}$ (deg)                   & $-9.9 \pm 1.5$\phs            & [$-90$, 90]    \\ [1ex]
\hline \\ [-1.5ex]
\multicolumn{3}{c}{Error Adjustment Parameters} \\ [0.5ex]
\hline \\ [-1.5ex]
 $f_{\rm RV1}$                           & $0.518 \pm 0.054$             & [$-5$, 5]      \\
 $f_{\rm RV2}$                           & $1.67 \pm 0.18$               & [$-5$, 5]      \\
 $f_{\rm P}$                             & $0.679 \pm 0.082$             & [$-5$, 5]      \\
 $f_{\rm Z}$                             & $0.86 \pm 0.14$               & [$-5$, 5]      \\
 $f_{\rm occ}$                           & $1.90 \pm 0.60$               & [$-5$, 5]      \\
 $f_{\rm Hip}$                           & $0.83 \pm 0.11$               & [$-5$, 5]      \\
 $\sigma_{V^2}$ for PAVO                 & $0.0469 \pm 0.0022$           & [0, 1]         \\
 $\sigma_{V^2}$ for CLIMB                & $0.0778 \pm 0.0088$           & [0, 1]         \\
 $\sigma_{\rm CP}$ for CLIMB (deg)       & $6.6 \pm 2.8$                 & [0, 30]        \\
 $\sigma_{\rm CP}$ for MIRC-X (deg)      & $1.882 \pm 0.045$             & [0, 30]        \\
 $\sigma_{\rm CP}$ for MYSTIC (deg)      & $0.495 \pm 0.053$             & [0, 30]        \\ [1ex]
\hline \\ [-1.5ex]
\multicolumn{3}{c}{Derived Properties} \\ [0.5ex]
\hline \\ [-1.5ex]
 $i_{\rm orb}$ (deg)                     & $107.863 \pm 0.032$\phn\phn   & \nodata        \\
 $e$                                     & $0.23565 \pm 0.00011$         & \nodata        \\
 $\omega_2$ (deg)                        & $335.697 \pm 0.082$\phn\phn   & \nodata        \\
 $a$ (au)                                & $1.768 \pm 0.018$             & \nodata        \\
 $M_1$ ($M_{\sun}$)                      & $5.04 \pm 0.17$               & \nodata        \\
 $M_2$ ($M_{\sun}$)                      & $3.64 \pm 0.12$               & \nodata        \\
 $q \equiv M_2/M_1$                      & $0.721 \pm 0.014$             & \nodata        \\
 $\pi_{\rm orb}$ (mas)                   & $7.340 \pm 0.076$             & \nodata        \\
 Distance (pc)                           & $136.2 \pm 1.4$\phn\phn       & \nodata        \\
 $\pi_{\rm Hip}$ (mas)                   & $8.08 \pm 0.46$               & \nodata        \\
 $(F_2/F_1)_{\rm Hip}$                   & $0.179 \pm 0.044$             & \nodata        \\
 $\mu_{\alpha}^*$ (mas yr$^{-1}$)        & $+18.21 \pm 0.44$\phn\phs     & \nodata        \\
 $\mu_{\delta}$ (mas yr$^{-1}$)          & $-44.63 \pm 0.35$\phn\phs     & \nodata        \\
 $\phi_{\rm maj}$ (mas)                  & $0.5340 \pm 0.0030$           & \nodata        \\
 $\phi_{\rm min}/\phi_{\rm maj}$         & $0.8456 \pm 0.0076$           & \nodata       
\enddata

\tablecomments{The values listed correspond to the mode of the
  posterior distributions, with uncertainties representing the 68.3\%
  credible intervals. Priors in square brackets are uniform over the
  ranges specified, except those for the error inflation factors $f$,
  which are log-uniform. The time of periastron passage, $T_{\rm peri}$,
  is referenced to BJD~2,400,000.}

\end{deluxetable}
\setlength{\tabcolsep}{6pt}

Despite the approximation of a circular orbit for Atlas used in the Hipparcos
catalog, the semimajor axis of the photocenter reported there ($4.23
\pm 0.97$~mas) is consistent with our more precise value in
Table~\ref{tab:mcmc}. Similarly for the Hipparcos inclination angle
($108\arcdeg \pm 25\arcdeg$). As expected, our revised Hipparcos
parallax for Atlas ($\pi_{\rm HIP} = 8.08 \pm 0.46$~mas) is about 0.7~mas larger
than the more precise orbital parallax, reflecting the known overestimate in
the satellite results for the Pleiades.

With our measurement of the semimajor axis of the photocenter from
the Hipparcos observations, it is straightforward to compute
the magnitude difference between the
components in the Hipparcos bandpass ($\Delta H\!p$).
It follows from the relation $a^{\prime\prime}_{\rm phot} =
a^{\prime\prime} (B - \beta)$, in which $B = M_2/(M_1+M_2)$ is the mass
fraction of the secondary, and $\beta$ is its fractional light
contribution, expressed also as $\beta = 1/(1+10^{0.4 \Delta
H\!p})$. We obtain $\Delta H\!p = 1.87 \pm 0.27$~mag.

\section{Discussion}
\label{sec:discussion}

\subsection{Properties of the Primary of Atlas}
\label{sec:primaryproperties}

Our discovery that the apparent disk of the primary of Atlas is
reasonably well represented by an ellipse, rather than a circle,
enables other properties of the rapidly rotating star to be inferred, with proper
consideration of projection effects. While the measured apparent major axis of
the ellipse, $\phi_{\rm maj}$, coincides with the equatorial
diameter of the star, the true polar diameter can only be determined
if we know the inclination angle $i$ of the rotation axis relative
to the line of sight. The apparent minor axis as measured, $\phi_{\rm min}$,
will generally be larger than the true polar diameter $\phi_{\rm
  pol}$ due to projection, and for an oblate ellipsoidal object,
  the two are related by
\begin{equation}
    \phi_{\rm min} = \sqrt{\phi_{\rm pol}^2\sin^2 i + \phi_{\rm maj}^2\cos^2 i}~.
    \label{eq:phimin}
\end{equation}
Then, for a given inclination angle, the linear polar radius of
the star, $R_{\rm pol}$, follows directly from $\phi_{\rm pol}$ and
the known distance. The inclination angle may be estimated by using
our measured projected rotational velocity of the primary, as follows.
Under the
assumption of hydrostatic equilibrium, uniform rotation, and a point
mass gravitational potential, the equatorial rotational velocity
of a star is given by
\begin{equation}
   v_{\rm eq} = \sqrt{ \frac{2GM}{R_{\rm pol}} \left(1 - \frac{R_{\rm
pol}}{R_{\rm eq}}\right)}
\label{eq:veq}
\end{equation}
\citep[e.g.,][]{Jones:2015}, in which $G$ is the gravitational
constant, $M$ is the stellar mass, and $R_{\rm pol}$ and $R_{\rm eq}$
are the polar and equatorial radii, respectively. Multiplying the expression above
by $\sin i$, and equating the right-hand side to our
measured $v \sin i$ value of 217~\kms, allows us to solve Eq.[\ref{eq:phimin}]
and Eq.[\ref{eq:veq}] jointly for the inclination angle and $R_{\rm pol}$.
We obtained $i = 64\arcdeg \pm 20\arcdeg$ (or $116\arcdeg \pm 20\arcdeg$) and
$R_{\rm pol} = 6.48 \pm 0.50~R_{\sun}$. Within its admittedly
large uncertainty, {the larger of the two values of $i$ is consistent
with being the same as the orbital inclination angle ($i_{\rm orb} = 107\fdg9$).
Furthermore, the nominal orientation of the primary's sky-projected disk
($\theta_{\phi} = -9\fdg9$, or $350\fdg1$) is also rather similar to the
position angle of the line of nodes for the orbit ($\Omega = 334\fdg2$),
although we note that $\theta_{\phi}$ has an inherent 180\arcdeg\
ambiguity that we cannot resolve, stemming from the cosine squared term in Eq.[\ref{eq:1}]. Nevertheless,
with the appropriate choices for both $i$ and $\theta_{\phi}$, we
conclude that the orbital and spin axes may well be close to alignment.
Indeed, proceeding with those choices and with the formal uncertainties for the
primary's disk properties, and keeping in mind the caveats mentioned earlier, we
obtain a true relative angle between those axes of
$\psi = 21\arcdeg \pm 12\arcdeg$.\footnote{This estimate of the true relative
inclination follows
from the expression $\cos \psi = \cos i_{\rm orb} \cos i +
\sin i_{\rm orb} \sin i \cos(\Omega - \theta_{\phi})$. We report
the median and standard deviation from a Monte Carlo exercise.}}

With the equatorial radius $R_{\rm eq} = 7.81 \pm 0.18~R_{\sun}$
(from $\phi_{\rm maj}$ and the distance), the
true oblateness is then $R_{\rm pol}/R_{\rm eq} = 0.828 \pm 0.057$. This is
a slightly smaller number (greater rotational flattening) than the apparent
oblateness of 0.846 reported above in Table~\ref{tab:mcmc}. The estimate
of $i$ implies a rotational velocity at the equator 
of $v_{\rm eq} = 233 \pm 45~\kms$. Given the simple nature of our
toy model for the primary, the uncertainties reported above have been
conservatively increased by a factor of two over their formal
values, which we consider to be too optimistic.

\subsection{The Distance}

\begin{figure}
\epsscale{1.17}
\plotone{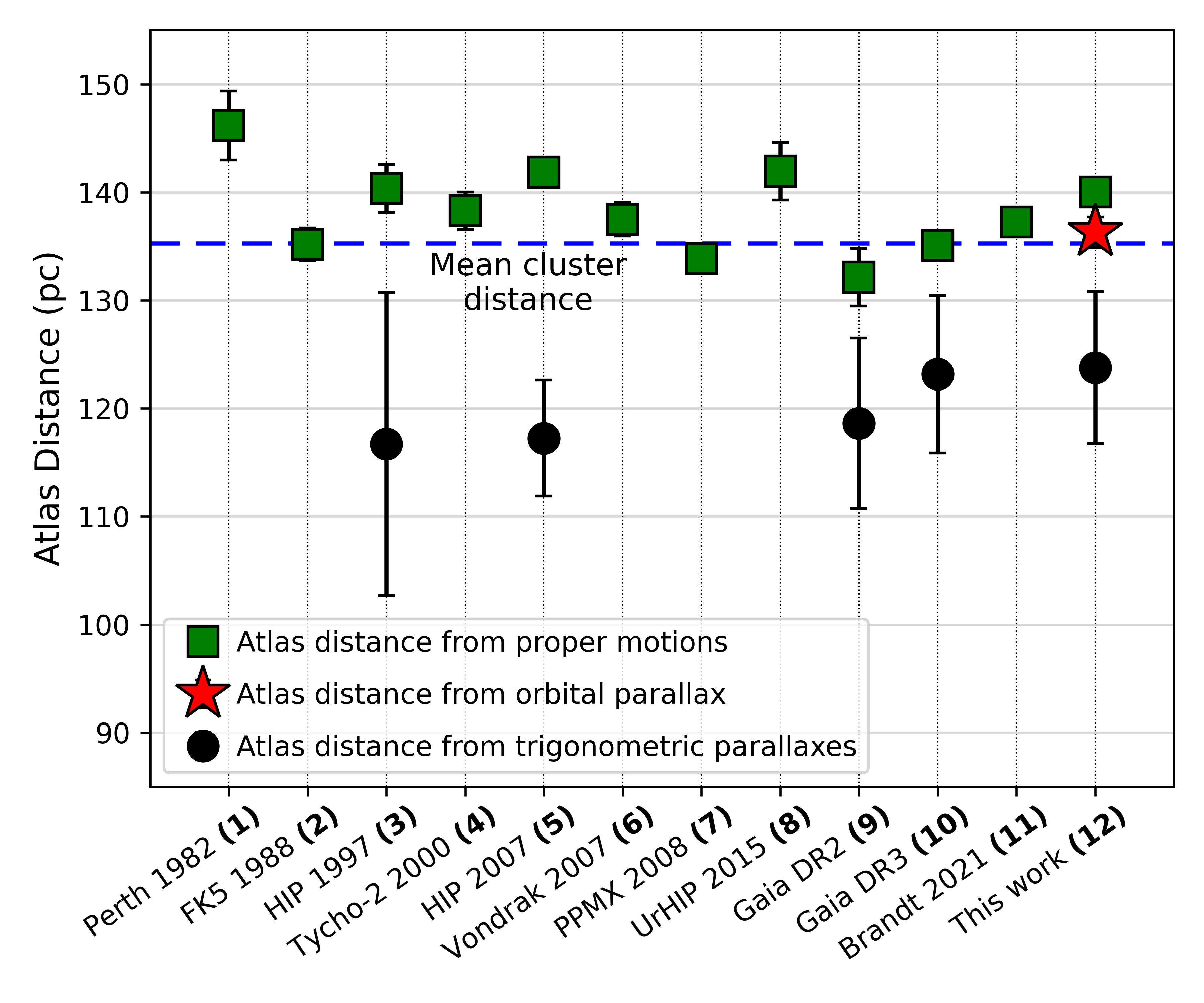}
\figcaption{Kinematic determinations of the distance to Atlas
(moving cluster method), based on its proper motion from various
sources, as labeled on the $x$ axis. Also indicated are
the few direct determinations from the trigonometric parallax,
as well as our orbital parallax result. The dashed line represents
the mean cluster distance from Gaia~DR2 \citep{Gaia:2018a}, after a
slight zeropoint adjustment following \cite{Lindegren:2018},
giving 135.3~pc.
Source numbers on the $x$ axis correspond to:
(1) \cite{Nikoloff:1982};
(2) \cite{Fricke:1988};
(3) \cite{ESA:1997};
(4) \cite{Hog:2000};
(5) \cite{vanLeeuwen:2007};
(6) \cite{Vondrak:2007};
(7) \cite{Roser:2008};
(8) \cite{Frouard:2015};
(9) \cite{GaiaDR2:2018};
(10) \cite{GaiaDR3:2023};
(11) \cite{Brandt:2021};
(12) This work: the distance from the proper motion and
from the direct parallax measurement both rely on our reanalysis
of the Hipparcos intermediate data (Section~\ref{sec:orbit}).
\label{fig:pmdist}}
\end{figure}

Trigonometric parallax measurements for Atlas have been reported in
the original and revised editions of the Hipparcos catalog, as well
as by the Gaia mission, in both its second and third data releases
(DR2 and DR3). However, neither of the Gaia values accounted for the
orbital motion of the binary. Our
orbital parallax therefore provides the most accurate measure of the
distance to date.  As a check, additional estimates may be obtained by the
classical moving cluster method, which assumes a common space motion
for all members of the Pleiades. These ``kinematic'' distances rely only on the known
space motion of the cluster, the sky position of the convergent point
and of Atlas itself, and the measured proper motion of the object.
Figure~\ref{fig:pmdist} shows several of these estimates, which we
calculated from proper motion measurements compiled from the
literature, selected to be as independent as
possible. We adopted the space motion and convergent point coordinates
for the cluster based on Gaia~DR2 \citep{GaiaDR2:2018}. Also
shown in the figure are the distances inferred from the available direct
trigonometric parallax measurements (Hipparcos, Gaia), which are all
too small.\footnote{As a demonstration of the bias for Gaia~DR3, we
carried out numerical simulations
following \cite{Perryman:2014}, and verified that not accounting for
orbital motion causes a systematic error in the derived trigonometric
parallax of Atlas (source identifier 66526127137440128) toward
larger values, i.e., toward shorter distances, just as observed, by about the amount
we see in this star.}  On the other hand, the kinematic distances are all seen to
be quite close to the mean value for the Pleiades (135.3~pc; Gaia~DR2). The
distance inferred from the orbital parallax in the present work,
$136.2\pm 1.4$~pc, is only 0.7\% larger. 
Given that the position of Atlas on the sky is only about
half a degree from the cluster center, we conclude that its location
in space must also be very near the center of the Pleiades. This is
not unexpected given the mass segregation known to exist in the Pleiades
\citep[e.g.,][]{Raboud:1998, Converse:2008, Alfonso:2024}, as Atlas
is one of the heavier members of the cluster.

\subsection{Comparison with Stellar Evolution Models}

The physical properties of Atlas derived in this work allow for a meaningful
comparison against current models of stellar evolution. However,
standard models are inadequate for
this case, because of the large rotational velocity of the primary.
Centrifugal forces in rapid rotators cause significant departures
from spherical symmetry, and give rise to gravity darkening. 
The structure, global properties, and evolution of stars change
significantly, in ways that standard, spherically symmetric models
cannot account for \citep[see, e.g.,][]{Meynet:2000, Maeder:2010}.
While a small number of grids of 
rotating models have been published, they typically offer only a few
fixed values for the mass and rotation rate, and many other parameters
are also held fixed.

Here we have used the Modules for Experiments in Stellar Astrophysics
code \citep[MESA;][and references therein]{Paxton:2019}
to generate suitable evolutionary tracks. This provides us with greater
flexibility for tuning various parameters
of interest, particularly the initial rotation rate
\citep[see also][]{Gossage:2019}\footnote{See
\url{https://doi.org/10.5281/zenodo.7796366}}.
Specifically, the version of MESA
we used here is release r22.11.1 (2022). We adopted the solar metallicity,
which is essentially the composition we derived spectroscopically
(see Table~\ref{tab:spectro}).
Recent versions of MESA that follow 
\cite{Paxton:2019} utilize a new method of accounting for centrifugal distortion
in stellar structure. The prior implementation followed the method of 
\cite{Endal:1976}, with numerically stable calculations valid for rotation 
rates up to roughly $\omega/\omega_{\rm crit} = 0.6$. Here, $\omega$ is
the angular rotation frequency, and $\omega_{\rm crit}$ is the angular
frequency at which the centrifugal force would match gravity at the stellar equator (``critical'' rotation).
The new implementation 
of centrifugal distortion improves on these calculations, with validity up to 
$\omega/\omega_{\rm crit} \approx 0.9$. This new implementation derives from 
analytical fits to the Roche potential of a single star as it becomes distorted 
via centrifugal forces \citep{Paxton:2019}. When comparing evolutionary calculations, it is important to bear in mind that other models such as PARSEC \citep{Nguyen:2022}, mentioned below, do not use the recent implementation brought by \cite{Paxton:2019} based on the Roche potential, and instead use an implementation of the method of \cite{Endal:1976}. Overall, stellar behavior near critical rotation rates remains an unresolved matter in need of observational constraint.

In addition to the individual masses, several other empirical
properties of both stars in Atlas are available for the comparison with models.
They include the spectroscopic effective temperatures and surface gravities,
as well as the absolute radii, $R_1 = 7.56 \pm 0.63~R_{\sun}$
and $R_2 = 3.25 \pm 0.29~R_{\sun}$. 
The radius of the rotationally distorted primary star, calculated here
from its mass and the spectroscopic
$\log g$, represents a mean value inherited from the nature of $\log g$,
which is itself an average over the visible disk of the star.
The radius for the secondary follows
from the value for the primary and the spectroscopic radius ratio
(Table~\ref{tab:spectro}).

Another mean value of $R_1$ that is
independent of $\log g$, and is more directly connected to the interferometry
and our model for the shape of the primary,
may be estimated from the polar and equatorial radii reported previously.
These rely on a different spectroscopic property ($v \sin i$) and on the distance.
With the same assumption as before that the star is
reasonably well represented by an oblate ellipsoid,
the volume-equivalent radius can be calculated as $R_{\rm 1,vol} = (R_{\rm eq}^2 R_{\rm pol})^{1/3}$. We obtain $R_{\rm 1,vol} = 7.32 \pm 0.26~R_{\sun}$. The corresponding volume-equivalent surface gravity is $\log g_{\rm 1,vol} = 3.414 \pm 0.025$.
Both of these estimates are consistent with those more closely
related to the spectroscopy through $\log g$.
Further constraints inferred from the distorted shape of the
primary are its true oblateness, and the expected surface rotational
velocity at the equator (see Section~\ref{sec:primaryproperties}).

In the presence of rapid rotation and the resulting gravity darkening,
observed global properties
such as the temperature or luminosity become dependent on the inclination
angle of the spin axis relative to the line of sight. For example, a star
viewed close to pole-on ($i \approx 0\arcdeg$) would present a hotter
disk-averaged temperature
and a higher luminosity than one viewed at higher inclination angles,
because the polar temperature is hotter.
MESA and other codes
typically report directional averages of those properties over the surface
of the star. In the case of the primary of Atlas, our
knowledge of the inclination angle of its rotation axis
allows us to infer ``projected'' properties from the models, by applying adjustments to the MESA predictions that depend on $i$ and
the rate of rotation, $\omega/\omega_{\rm crit}$. These projected properties
will then more closely correspond to those we actually measured for the star.
Here we adopted corrections
based on the gravity-darkening model of \cite{EspinosaLara:2011}, as implemented
in the {\tt GDit}\footnote{\url{https://github.com/aarondotter/GDit/tree/master}} code written by Aaron Dotter.

Comparisons of the observations for the more evolved primary
star were made against evolutionary tracks from MESA,
adjusted as described above, for a wide
range of initial angular rotation rates relative to the breakup rate,
$\omega_0/\omega_{\rm crit}$. The prescription we used for
convective core overshooting is the diffusive approximation
\citep{Freytag:1996, Herwig:1997}, with an overshooting parameter
$f_{\rm ov} = 0.016$, identical to that adopted
in other sets of MESA-based models, such as the MESA Isochrones
and Stellar Tracks series \citep[MIST;][]{Choi:2016}. 
This $f_{\rm ov}$ value is also consistent with semi-empirical
estimates for slightly less massive stars in eclipsing binaries,
as reported by \cite{Claret:2019}.
The mixing length parameter was set to $\alpha_{\rm ML} = 1.82$, as
in the MIST calculations.

\begin{figure*}
\epsscale{1.17}
\plotone{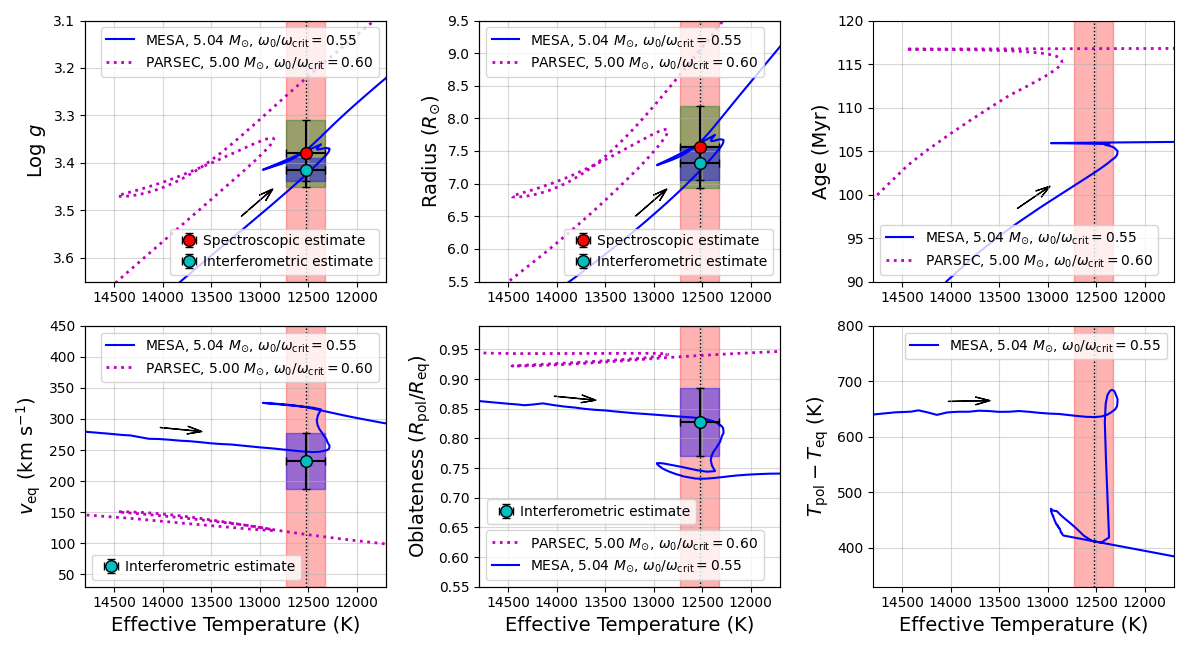}
\figcaption{Comparison of the measured properties of the primary
of Atlas against theory. The blue line corresponds to our MESA
model for the nominal mass of $M_1 = 5.04~M_{\sun}$ determined in this work,
Solar metallicity was assumed. The initial rotation rate was set
to $\omega_0/\omega_{\rm crit} = 0.55$, and the convective overshooting
parameter was fixed at $f_{\rm ov} = 0.016$. The dotted magenta line is a model
from the PARSEC~v2.0 series, with properties similar to those above. 
Both evolutionary tracks have been adjusted for the effects
of gravity darkening, as described in the text.
Arrows indicate the direction of evolution.
The measured temperature is represented by the
shaded area, and $R_1$, $\log g$, $v_{\rm eq}$, and the oblateness
are shown with their corresponding error boxes. The two values shown
for the radius and $\log g$ (`spectroscopic' and `interferometric')
were derived in different ways (see the text), but are consistent. \label{fig:mesa1}}
\end{figure*}

Extensive tests indicated that, within the uncertainties,
a model for a star spinning at an initial
rate of about 55\% of the breakup value ($\omega_0/\omega_{\rm crit} \approx 0.55$)
reaches a satisfactory agreement with the spectroscopically measured
$T_{\rm eff}$ and our two estimates of $\log g$, for the
nominal primary mass of $5.04~M_{\sun}$
(see Figure~\ref{fig:mesa1}, top left panel).
For comparison, we show also
an evolutionary track from the PARSEC~v2.0 series \citep{Nguyen:2022},
for a mass of 5.00~$M_{\sun}$ and $\omega_0/\omega_{\rm crit} = 0.60$,
which are the nearest available values to those adopted for our MESA
calculations. As seen in
the figure, the PARSEC model is some 500~K hotter than the one from MESA,
and therefore does not agree as well with the measurements for Atlas.

In Figure~\ref{fig:mesa1}, 
the evolutionary tracks suggest that the primary of Atlas is at or near 
the so-called blue hook, referring to the brief blueward ``hook'' 
exhibited by the models. This evolutionary phase typically follows core 
hydrogen exhaustion, and is due to a relatively brief halt in nuclear burning 
and subsequent contraction of the star, until hydrogen reignites in a shell 
around the inert helium core. Our MESA model displays some slight wiggles 
in its evolutionary track around this time that are associated with this 
contraction.
As the model contracts, its rotation rate increases and the model approaches a critical rotation rate at this point in the evolution. Prior to this contraction, the point at which the model most closely matches the temperature of the primary of Atlas corresponds to a rapid rotation rate of $\omega/\omega_{\rm crit} \approx 0.77$.
Consequently, theory predicts that the primary of Atlas may have a 
significantly distorted stellar structure under the effects of rapid rotation,
as we actually observe.

The radius vs.\ temperature plot is shown in the top middle panel, where
the match is similar to the previous panel, as expected given that $R_1$
depends on $\log g$.
The age of Atlas, as inferred from the MESA model, is between 102 and 106~Myr
(top right panel), which is not far from other age estimates
for the Pleiades using a variety of methods:
112~Myr \citep{Dahm:2015},
104--117~Myr \citep{Naylor:2009},
118~Myr \citep{Frasca:2025},
130~Myr \citep{Barrado:2004}, and
110--160~Myr \citep{Gossage:2018},
among many others.

The MESA and PARSEC models provide predictions for the change in the
equatorial rotational velocity
($v_{\rm eq}$) as a function of age. Within the uncertainties,
the expectation from MESA is consistent with our estimate of that
quantity described in Section~\ref{sec:primaryproperties}
($v_{\rm eq} \sim233~\kms$), while the PARSEC model underestimates it.
The middle panel at the bottom of Figure~\ref{fig:mesa1} displays
our oblateness estimate for the primary, along with the evolution of
this quantity expected from both models. Again, there is good
consistency between theory and the observation for MESA, assuming the
star is near the point of central hydrogen exhaustion, as suggested
by the other comparisons. On the other hand, the PARSEC model would
predict the star to have a more spherical shape than we measure. In order
to match our estimates of $v_{\rm eq}$ and the oblateness, the PARSEC
model would require a much higher initial rotation rate of
$\omega_0/\omega_{\rm crit} \sim 0.80$, but we find that
such a model would overestimate the radius, and underestimate $\log g$.

Convective core overshooting and rotation have somewhat similar
effects on the models, in the sense that they both favor
mixing of fresh hydrogen fuel into the core. This typically leads to an extension of the evolutionary tracks toward
cooler temperatures and higher luminosities, and results in longer main-sequence
lifetimes. One may therefore expect it might be possible for the
models to match the observations with a range of different combinations of
the strength of overshooting ($f_{\rm ov}$) and initial rotation
($\omega_0/\omega_{\rm crit}$). However, for Atlas we find that
our estimates of the current shape and equatorial rotation of the primary tend to
lift that degeneracy. They constrain the initial rotation to be near the value we report, given the measured mass and effective temperature of the star. Consequently, overshooting strengths much different from what we assumed do not improve the fit to the measurements.

Finally, in the bottom right panel of Figure~\ref{fig:mesa1} we include the predictions
from MESA for the difference between the polar and equatorial temperature
caused by gravity darkening, which our observations do not constrain. It
is expected to be roughly 650~K at the current evolutionary state of the star.

\begin{figure*}
\epsscale{1.17}
\plotone{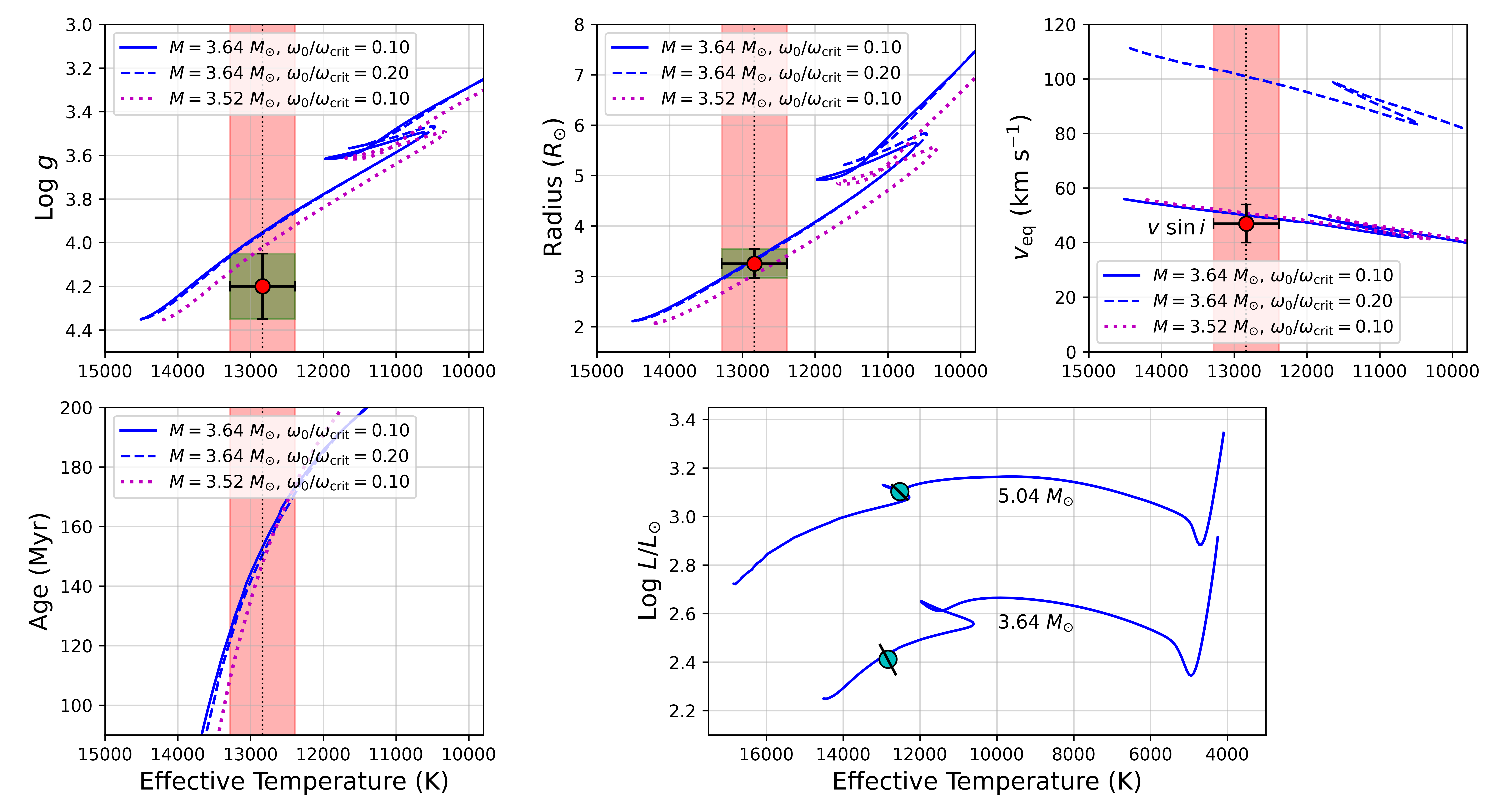}
\figcaption{Similar to Figure~\ref{fig:mesa1}, now comparing the spectroscopic
properties for the secondary of Atlas against theory, at its nominal mass of $M_2 = 3.64~M_{\sun}$. In this case, MESA models are shown for two different rotation rates, as labeled. An additional model for $M_2 = 3.52~M_{\sun}$ (1$\sigma$ lower than the nominal mass; dotted magenta line) provides a marginally better match to the measured $\log g$. The panel on the lower right illustrates the relative evolutionary states of the two components of Atlas. Stellar luminosities for this illustration were calculated from the effective temperatures and radii. {The error bars are tilted because of
the correlation between $L$ and $T_{\rm eff}$.} \label{fig:mesa2}}
\end{figure*}

Figure~\ref{fig:mesa2} presents an analogous comparison of the
observations for the secondary of Atlas against MESA models, for
the nominal mass of $M_2 = 3.64~M_{\sun}$.
Two different initial rotation rates are shown
($\omega_0/\omega_{\rm crit} = 0.10$ and 0.20), although their impact on the
surface gravity and radius as a function of effective temperature
is minimal.
We find only marginal agreement with the model in the $\log g$
vs.\ $T_{\rm eff}$ plane, but good agreement for the radius. 

It is quite possible that some of this disagreement has to do with the chemical
anomalies in the secondary.
While we did account for the underabundance of He and the overabundance of Fe, Cr, and Ti in the calculation of model atmospheres for the star, potential anomalies in the abundances of other elements were not investigated here.
Moreover, chemically peculiar stars are also known to exhibit vertical stratification of elements in their atmospheres, which influences the profiles of physical quantities such as temperature and pressure — ultimately leading to further perturbations in the atmospheric structure \citep[e.g.,][]{Shulyak2009,Pandey2011,Makaganiuk2012}. These perturbations were not taken into account in our spectroscopic analysis.
Additionally, the method of spectral disentangling used in this study assumes that the line profiles do not change shape as a function of time — an assumption that is violated in the case of the secondary. Given these issues, we cannot rule out potential biases in the atmospheric properties for the secondary that are not reflected in the purely statistical errors reported in this work.

Because the secondary component is relatively unevolved, the
age is not well constrained by the observations (i.e., the tracks are nearly vertical
in the bottom left panel of Figure~\ref{fig:mesa2}), and our estimate is highly sensitive to the temperature. With these models, the secondary appears slightly
older than the primary. This may be related as well to the anomalous composition
of the star, which is not taken into account in the MESA models.
A somewhat better match to $\log g$ can be achieved by adopting
a model mass 1$\sigma$ lower than the nominal value ($M_2 = 3.52~M_{\sun}$;
see Figure~\ref{fig:mesa2}, dotted line).
This also yields better agreement with the age inferred for the primary.

The rotational velocity comparison is shown in the top right panel of {Figure~\ref{fig:mesa2}}.
As the inclination angle of the secondary's spin axis is unknown
(but see below),
here we have compared the predicted true equatorial velocity from theory
against our measured \emph{projected} equatorial velocity
($v \sin i = 47~\kms$).
There is formal agreement with the model for $\omega_0/\omega_{\rm crit} = 0.10$,
within the uncertainty, but any inclination angle different from 90\arcdeg\
would convert the empirical measurement to a higher equatorial value,
and imply a true initial angular rotation rate larger than this.

Finally, the lower right panel of Figure~\ref{fig:mesa2} displays the
primary and secondary of Atlas together in the H-R diagram, to illustrate
their relative states of evolution.

\section{Final Remarks}
\label{sec:conclusions}

With the present analysis, Atlas now ranks among the better
characterized binary systems in the Pleiades cluster, both in terms of
its improved (3D) orbit and the physical properties of the individual
components. Our spectroscopic and interferometric observations have
revealed two salient features of the stars in the system. One is that
the primary is a rotationally distorted object, for which we have been
able to measure the oblateness ($R_{\rm pol}/R_{\rm eq} \approx 0.83$)
directly from the interferometric observations. We have also obtained
a rough estimate of
the {orientation} of its spin axis, which is possibly
aligned with the axis of the orbit. Only about a dozen rapidly
rotating, early-type stars have had their distended shapes determined
in this way, as instrumental limitations of the interferometric technique
typically require them to be
bright and nearby. Atlas is by far the most distant example, all
others being closer than 50~pc. Additional interferometric
observations to support a more sophisticated analysis that accounts
for gravity darkening and other effects should be able to improve upon
our estimates.  A second result of interest is that while the chemical
composition of the primary appears to be essentially solar, consistent
with the known metallicity of the cluster, the secondary is a
helium-weak star with significant enhancement of Fe, Cr, and
Ti, and perhaps other elements. The latter three are about 10 times
more abundant than in the Sun, on average, while He is about 8 times
weaker.

Atlas had previously been classified as a He-weak object in the
catalog of \cite{Renson:2009}, but until now it had not been
established which of the two components has the anomaly, or whether
both do. Our spectroscopic analysis has now revealed the secondary to
be the culprit. Together with its other abundance abnormalities, which are found
with similar patterns in about 10\% of all A- and B-type main-sequence
stars, this places it in the ``ApBp'' class of chemically peculiar objects.
Stars in this class are often found to also have strong magnetic
fields \citep[e.g.,][]{Smith:1996, Briquet:2007, Petit:2025}, which
are considered to be of fossil origin, i.e., descended from the fields
present in the natal molecular clouds. In fact, \cite{Neiner:2015}
used the Zeeman signature to determine that the secondary of Atlas has
a dipolar magnetic field with a strength of several hundred Gauss. The
primary, on the other hand, shows no indication of having one. From
changes in the field strength on different nights, they concluded that
the magnetic axis of the secondary is not aligned with its axis of
rotation, which is not uncommon.  Abundance anomalies in these
magnetic, early-type stars have also been associated with the presence
of chemical patches or ``spots'' on the surface \citep[e.g.,][]{David-Uraz:2019},
which naturally lead to line profile variations as the star rotates or
the spots change. We observe such variations very clearly in the
secondary.

Several authors have established that Atlas is photometrically
variable \citep[e.g.,][]{McNamara:1985, McNamara:1987, Wraight:2012, Zwintz:2024}.
The detailed study by \cite{White:2017} identified several
frequencies corresponding to periods of $\sim$2.5 days or less, some of
which are the same as seen by others. The frequency with the largest
amplitude (2.1~mmag) corresponds to a period of 2.428~d.
\cite{White:2017} concluded that these
frequencies, if they all originate on the same star,
are most likely due to pulsations, rather than rotation,
and argued that they probably come from the primary, which is more
than five times brighter than the secondary. Nevertheless, they
cautioned that further investigation is required for confirmation.
However, if the 2.428~d periodicity were
in fact due to rotation,
it could not be caused by the primary, as our
measured radius and $v \sin i$ for that star would imply
$v_{\rm eq} < v \sin i$. On the other hand, it
could well come from the secondary, and in that case, the inferred
inclination angle of its spin axis would be $44\arcdeg \pm 10\arcdeg$.

The distance to Atlas is now known to 1\%, and the masses of the
components to about 3\%, currently limited by the spectroscopy. Other
relevant properties determined here for both stars include the
effective temperatures, radii, $\log g$, and the projected rotational
velocities.  Atlas is one of only four binaries in the Pleiades
cluster that have dynamical mass determinations to date, and is the
most massive. The other three are
HII~1431 \citep[HD~23642;][]{Munari:2004, Southworth:2005, Groenewegen:2007, David:2016, Southworth:2023}, HCG~76
\citep[V612~Tau;][]{David:2016}, and HII~2147 \citep{Torres:2020}.
The measured properties of the primary of Atlas are found to be in
good agreement with stellar evolution models from MESA that account
for the rapid rotation of the star, and suggest the initial rotation
rate on the zero-age main sequence was about 55\% of the breakup
velocity. The current rotation rate is estimated to be
$\omega/\omega_{\rm crit} \approx 77$\%, according to these models.
The comparison with theory places the star at the very end of the main
sequence, prior to core hydrogen exhaustion (the beginning of the blue
hook), at an age between 102 and 106~Myr that is similar to other estimates for
the cluster. The secondary component rotates more slowly, and the
predictions from MESA models are also generally consistent with its
other properties at its measured mass.

\begin{acknowledgements}

The spectroscopic observations of Atlas were gathered with the
assistance of P.\ Berlind, M.\ Calkins, G.\ Esquerdo, and A.\ Medina.
J.\ Mink (CfA) is thanked for maintaining the database of \'echelle observations
at the CfA. We are also grateful to R.\ Matson (USNO) for providing a record of
the astrometric observations of Atlas from the Washington Double Star
Catalog, maintained at the U.S.\ Naval Observatory (USNO),
and to J.\ Jones
(Georgia State Univ.) for information about systematic uncertainties
for the PAVO beam combiner.
{S.\ Chhabra,
I.\ Codron,
C.\ L.\ Davies,
J.\ Ennis,
T.\ Gardner,
M.\ Gutierrez,
D.\ Huber,
N.\ Ibrahim,
J.-B.\ Le Bouquin,
V.\ Maestro,
B.\ R.\ Setterholm,
T.\ ten Brummelaar, and
P.\ Tuthill
contributed to the CHARA observations of Atlas in other ways,
either as instrument PIs, members of the MIRC-X/MYSTIC development
and commissioning teams, or as observers. We thank them all.
We also acknowledge helpful comments by the anonymous referee.}

This work is based upon observations obtained with the Georgia State University Center for High Angular Resolution Astronomy Array at Mount Wilson Observatory.  The CHARA Array is supported by the National Science Foundation under Grant No.\ AST-1636624, AST-2034336, and AST-2407956. Institutional support has been provided from the GSU College of Arts and Sciences and the GSU Office of the Vice President for Research and Economic Development. Time at the CHARA Array was granted internally and through the NOIRLab community access program (NOIRLab PropID: 2018B-0326; PI: C.\ Melis; NOIRLab PropID: 2022B-235883; PI: G.\ Torres). SK acknowledges funding for MIRC-X received from the European Research Council (ERC) under the European Union's Horizon 2020 research and innovation program (Starting Grant No.\ 639889 and Consolidated Grant No.\ 101003096). JDM acknowledges funding for the development of MIRC-X (NASA-XRP NNX16AD43G, NSF-AST 2009489) and MYSTIC (NSF-ATI 1506540, NSF-AST 1909165). 
SJM was supported by the Australian Research Council through Future Fellowship FT210100485. SG acknowledges support from the Gordon and Betty Moore Foundation under project numbers GBMF8477 and GBMF12341.
This research has made use of the Jean-Marie Mariotti Center (JMMC) Aspro and SearchCal services.

This work has also made use of the SIMBAD and VizieR databases, operated at
the CDS, Strasbourg, France, and of NASA's Astrophysics Data System
Abstract Service. The computational resources used for this research
include the Smithsonian High Performance Cluster (SI/HPC), Smithsonian
Institution (\url{https://doi.org/10.25572/SIHPC}).

\end{acknowledgements}


\end{document}